\documentclass[reprint, superscriptaddress, amsmath,amssymb, aps, longbibliography]{revtex4-1}
\usepackage{graphicx}
\usepackage{dcolumn}
\usepackage{bm}
\usepackage{nicefrac,xfrac}
\usepackage{hyperref}
\begin{document}
\title{Optimal broad-band frequency conversion via a magnomechanical transducer}
\author{F. Engelhardt}
\email{fabian.engelhardt@mpl.mpg.de}
\affiliation{Max Planck Institute for the Science of Light, Staudtstr. 2, PLZ 91058 Erlangen, Germany}
\affiliation{Department of Physics, University Erlangen-Nuremberg, Staudtstr. 7, PLZ 91058 Erlangen, Germany} 
\affiliation{Institute for Theoretical Solid State Physics, RWTH Aachen University, 52074
Aachen, Germany} 
\author{V. A. S. V. Bittencourt}
\affiliation{Max Planck Institute for the Science of Light, Staudtstr. 2, PLZ 91058 Erlangen, Germany}
\author{H. Huebl}
\affiliation{Walther-Mei\ss ner-Institut, Bayerische Akademie der Wissenschaften, D-85748 Garching, Germany}
\affiliation{Physik-Department, Technische Universit\"{a}t M\"{u}nchen, D-85748 Garching, Germany}
\affiliation{Munich Center for Quantum Science and Technology (MCQST), D-80799 Munich, Germany}
\author{O. Klein}
\affiliation{Universit\'{e} Grenoble Alpes, CEA, CNRS, Grenoble INP, Spintec, 38054 Grenoble, France}
\author{S. Viola Kusminskiy}
\email{silvia.viola-kusminskiy@mpl.mpg.de}
\affiliation{Institute for Theoretical Solid State Physics, RWTH Aachen University, 52074
Aachen, Germany}
\affiliation{Max Planck Institute for the Science of Light, Staudtstr. 2, PLZ 91058 Erlangen, Germany}
\date{September 26 2022}
\begin{abstract}
Developing schemes for efficient and broad-band frequency conversion of quantum signals is an ongoing challenge in the field of modern quantum information. Especially the coherent conversion between microwave and optical signals is an important milestone towards long-distance quantum communication. In this work, we propose a two-stage conversion protocol, employing a resonant interaction between magnetic and mechanical excitations as a mediator between microwave and optical photons. Based on estimates for the coupling strengths under optimized conditions for yttrium iron garnet, we predict close to unity conversion efficiency without the requirement of matching cooperativities. We predict a conversion bandwidth in the regions of largest efficiency on the order of magnitude of the coupling strengths which can be further increased at the expense of reduced conversion efficiency.
\end{abstract}
\maketitle
\section{Introduction}
Modern quantum information technology relies on quantum control across several platforms, ranging from single atoms, ions and spins, to superconducting circuits. A given system is, however, usually only able to excel at a specific task.
For example, superconducting qubits and superconducting cavities, which typically operate in the microwave energy range, have been identified as one of the front-runners \cite{aruteQuantumSupremacyUsing2019} for quantum information processing, whereas transferring quantum information over long distances \cite{magnardMicrowaveQuantumLink2020} is a task best accomplished with optical signals travelling inside photonic fibers  \cite{xavierQuantumInformationProcessing2020}. Critically coupling microwave quantum circuits with optical communication networks represents a step forward in the development of a quantum network \cite{lodahlQuantumdotBasedPhotonic2018, kozlowskiLargeScaleQuantumNetworks2019} and requires the design of microwave to optics conversion setups with high efficiency and broad bandwidth.\par
Direct conversion between optical and microwave frequencies is possible in materials with significant non-linearities in their electric polarization \cite{boydNonlinearOptics2008, tsangCavityQuantumElectrooptics2011a, savchenkovParametricOscillationsWhispering2007, khanOpticalDetectionTerahertz2007, strekalovEfficientUpconversionSubterahertz2009, ruedaElectroopticEntanglementSource2019,fanSuperconductingCavityElectrooptics2018a, heaseBidirectionalElectroOpticWavelength2020}. In essence, large second order non-linearities give rise to three-wave interactions, such as parametric down-conversion and Stokes and anti-Stokes processes. Those have been used to achieve conversion efficiencies on the order of $10^{-1}$ at a bandwidth of $\approx18\,\rm{MHz}$ \cite{sahuQuantumenabledOperationMicrowaveoptical2022}.
Instead of direct conversion (zero-stage conversion), one can use one or multiple excitations as mediating elements. In state-of-the-art conversion platforms, mostly one-stage conversion setups are considered and explored, with a single mediator between optics and microwaves. For example, mechanical resonators can be used to couple vibrational modes to microwaves and optics simultaneously via electromechanical \cite{blencoweQuantumElectromechanicalSystems2004, lyshevskiElectromechanicalSystemsDevices2011} and optomechanical interactions \cite{regalCavityElectromechanicsCavity2011, parkinsQuantumStateTransfer1999, safavi-naeiniProposalOptomechanicalTraveling2011, andrewsBidirectionalEfficientConversion2014a} respectively. The latter is achieved by confining an optical mode in the mechanical resonator itself, which has been realized, for example, in optomechanical crystals \cite{aspelmeyerCavityOptomechanicsNano2014a, eichenfieldOptomechanicalCrystals2009a, chanOptimizedOptomechanicalCrystal2012b}. In experiments using a dielectric membrane conversion efficiencies of $\xi \sim 10^{-1}$ was achieved \cite{higginbothamHarnessingElectroopticCorrelations2018a} at a bandwidth of $\Delta \approx100\,\rm{kHz}$, which is limited by the total linewidth of the mechanical mode. \par
Beyond mechanical excitations, collective magnetic excitations, called magnons, have become promising candidates for quantum information processing \cite{chumakMagnonSpintronics2015a, lachance-quirionHybridQuantumSystems2019a, chumakAdvancesMagneticsRoadmap2022} and, in particular, for frequency conversion. Magnons allow frequency tunability through external magnetic fields and can be coupled strongly to microwaves via magnetic dipole interaction \cite{zhangBroadbandNonreciprocityEnabled2020a, hueblHighCooperativityCoupled2013b, pottsStrongMagnonPhoton2020}, which can be used to mediate the coupling between magnons and superconducting qubits \cite{tabuchiQuantumMagnonicsMagnon2016}. Additionally, magnetization and light couple through magneto-optical effects \cite{landauElectrodynamicsContinuousMedia2013a}, e.g., the Faraday effect, which however are intrinsically weak. Confining light inside the magnet enhances the magnon-photon coupling \cite{violakusminskiyCoupledSpinlightDynamics2016b, sharmaOptimalModeMatching2019}, even though achieving sufficient overlap between the modes is challenging \cite{grafCavityOptomagnonicsMagnetic2018a, grafDesignOptomagnonicCrystal2021a}. The weakness of magneto-optical effects together with the challenge of achieving good mode matching strongly limits magnon-based microwave-to-optics transduction in current experiments \cite{hisatomiBidirectionalConversionMicrowave2016a, haighTripleResonantBrillouinLight2016a}. Presently, state-of-the-art efficiencies of $10^{-7}$ \cite{zhuWaveguideCavityOptomagnonics2020a} at a bandwidth of $\approx16\,\rm{MHz}$ have been demonstrated. Nevertheless, magnetostriction, which couples the magnetization with mechanical deformations on a static and dynamic level in a material, is a relevant and even dominant interaction in magnetic materials \cite{spencerMagnetoacousticResonanceYttrium1958a, shengSpinWavePropagation2020, schwienbacherMagnetoelasticityCo25Fe75Thin2019, weilerVoltageControlledInversion2009}. Recent demonstrations of large coherent interaction between magnons and phonons \cite{anCoherentLongrangeTransfer2020, anBrightDarkStates2022, schlitzMagnetizationDynamicsAffected2022a, hatanakaOnChipCoherentTransduction2022} provides an incentive to interface optomechanics with magnomechanics in a hybrid system tailored for optics-to-microwave multi-stage transduction. \par
In this article, we propose an approach to optics-to-microwave conversion based on optomechanics by inserting a magnetic element that maintains high cooperativity both with a mechanical mode and with the microwave mode. The cost of the increased setup complexity comes at the gain of substantial increase of the coupling strength: through magnetoelastic and optomechanical effects it is possible to achieve a much stronger coupling to the microwave mode, compared to purely mechanical systems, and to optical photons, compared to the purely magnetic system. 
We show that such a optomagnomechanical setup, schematized in Fig. \ref{fig:model_interactionscheme}, yields conversion efficiencies close to unity, while simultaneously providing large bandwidths. The latter is due to the fact that we expect all relevant coupling strengths to be large compared to any internal linewidths. Our estimates are based on the magnetic insulator yttrium iron garnet (YIG). This material is the staple choice for magnonic experiments, since it provides long lifetime of magnon excitations and a large spin density \cite{mallmannYttriumIronGarnet2013, klinglerGilbertDampingMagnetostatic2017, maier-flaigTemperaturedependentMagneticDamping2017, kosenMicrowaveMagnonDamping2019}. The strong coupling of the considered modes leads to hybridization and the formation of frequency-shifted normal modes. Whereas the phonon frequencies are mainly determined by the choice of a specific geometry, magnons provide a desirable tunability for optimization. When all modes are tuned to resonance, the efficiency is given only in terms of the cooperativities between the modes. The cooperativity relates the coupling strength between modes to their linewidths, therefore characterizing the dissipative behaviour of a coupled system. Contrary to setups based on one-stage conversion, matching cooperativities are not required in order to obtain efficiencies close to unity. Here, optimization can be performed by altering the optical pump power (which determines the effective optomechanical coupling rate) and the respective port-coupling rates of the optical and microwave mode. Moreover, the total linewidth of the optical modes determines the optomechanical coupling regime and therefore the hybridization of the coupled system. \par
The article is structured as follows. In Section \ref{section:model} we introduce the model Hamiltonian and give a brief description of each relevant coupling term together with an estimate of the coupling strength for material-specific parameters for YIG. These values together with relevant frequencies and linewidths are summarized in Tab. \ref{tab:values}. In Section \ref{section:protocol} we set up the two-stage conversion setup based on resonant magnon-phonon coupling and calculate the expected conversion efficiency. We give a description of the expected hybridization of the interaction chain and discuss the resulting bandwidths of the conversion spectrum for different regimes of the driven optomechanical interaction. Finally, in Section \ref{section:outlook} we summarize our results.
\section{Model \label{section:model}}
\begin{figure}
    \centering
    \includegraphics[width=0.48\textwidth]{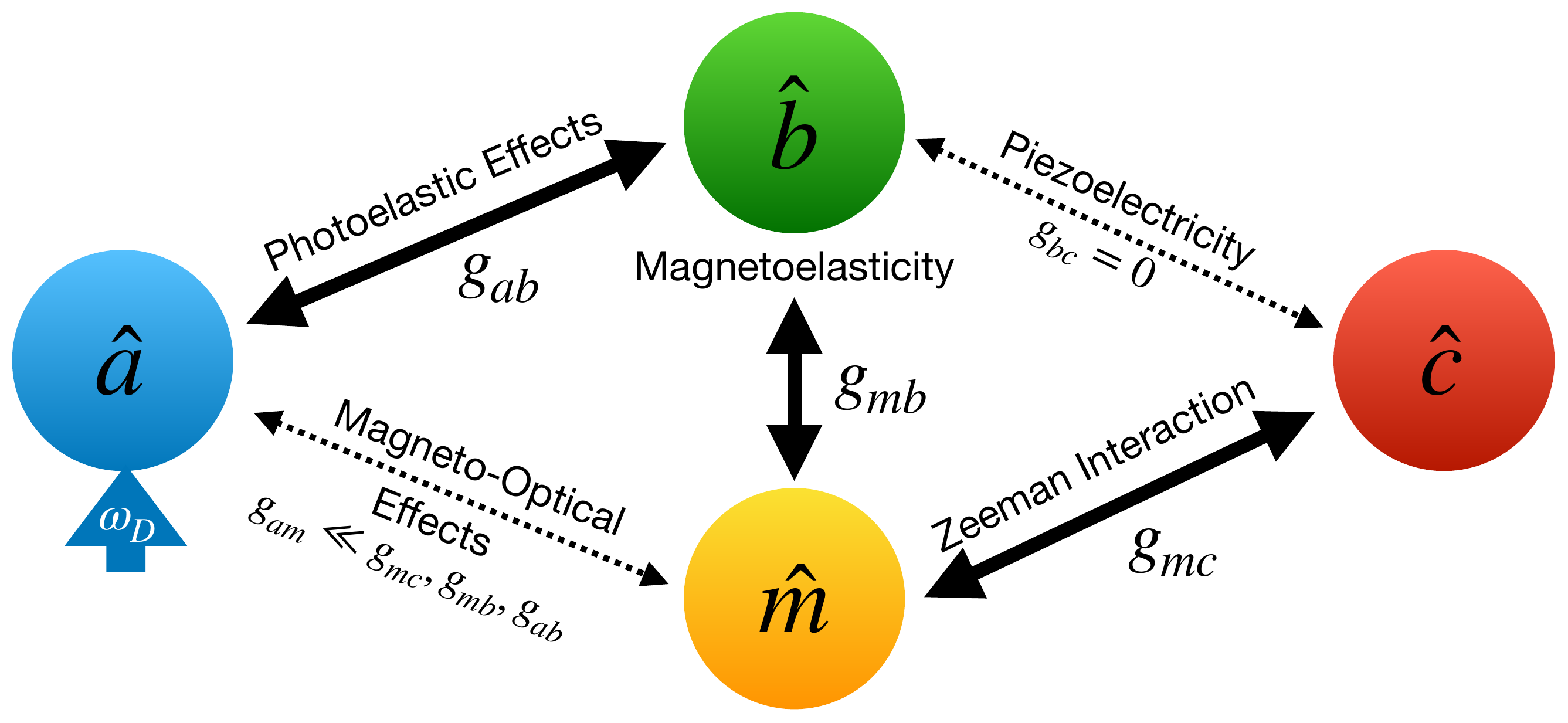}
    \caption{Sketch of a two-stage interconversion protocol between optical and microwave photons represented respectively by $\hat{a}$ and $\hat{c}$. The two intermediary states are $\hat{b}$, phonons and $\hat{m}$, the magnons, both with frequencies in the microwave range. The thick lines denote the dominant interaction scheme.}
    \label{fig:model_interactionscheme}
\end{figure}
Our model is based on the interaction scheme given in Fig. \ref{fig:model_interactionscheme}. Four different bosonic modes are coupled, with frequencies $\omega_a$, $\omega_b$, $\omega_m$, $\omega_c$ (where $\omega_a$ $\gg$ $\omega_b$, $\omega_m$, $\omega_c$) corresponding to optical photons, phonons, magnons and microwave photons, respectively. Optical photons couple to phonons and magnons via optomechanical and magneto-optical effects, respectively. Due to a large frequency mismatch, these interactions are parametric. Magnons couple to phonons and microwave photons via magnetoelastic and magnetic dipole interactions, respectively. We consider a setup in which magnons, phonons and microwave photons can be all three brought into degeneracy, leading to linear interaction terms. In this work, we consider no piezoelectric interactions and assume that the magneto-optical coupling is weak compared to all other types of interactions. This leads to an effective interaction chain comprising optomechanics, magnomechanics and inductive coupling to magnetization dynamics. \par
The undriven Hamiltonian of the system is given by
\begin{eqnarray}
   \hat{H}_1/ \hbar = && \omega_a \hat{a}^{\dagger}\hat{a} + \omega_m \hat{m}^{\dagger}\hat{m} + \omega_b \hat{b}^{\dagger}\hat{b} + \omega_c \hat{c}^{\dagger}\hat{c} \nonumber\\
    && g_{ab} \,\hat{a}^{\dagger}\hat{a}\left(\hat{b}^{\dagger} + \hat{b}\right) \nonumber
    + g_{mb} \left( \hat{b}^{\dagger} + \hat{b}\right)\left(\hat{m}^{\dagger} + \hat{m}\right)\\
    &&+ g_{mc} \left( \hat{m}^{\dagger} + \hat{m}\right)\left(\hat{c}^{\dagger} + \hat{c}\right).
    \label{eq:model_HamiltonianFull}
\end{eqnarray}
Here, $\hat{a}\left(\hat{a}^{\dagger}\right)$, $\hat{b}\left(\hat{b}^{\dagger}\right)$, $\hat{m}\left(\hat{m}^{\dagger}\right)$, $\hat{c}\left(\hat{c}^{\dagger}\right)$ are the bosonic annihilation (creation) operators for optical photon, phonon, magnon and microwave photons, respectively. Moreover, $g_{ab}$, $g_{mb}$ and $g_{mc}$ are the optomechanical, magnomechanical and inductive coupling strength, respectively. In order to realize the Hamiltonian of Eq. \eqref{eq:model_HamiltonianFull}, a microstructure serving as an optical cavity and at the same time supporting mechanical and magnetic modes in the GHz regime is needed. In the following we briefly describe the coupling mechanisms subject to our model and give estimates for the optimal expected couplings using material specific constants for YIG. Full derivations of the coupling terms can be found in the Appendix, as well as a summary of constants in Tab. \ref{tab:values_mat}. \par
The coupling between magnons and phonons is mediated by magnetoelastic effects, where the magnetization inside a material couples dynamically to the mechanical stress and thus causing lattice deformations. State-of-the-art experiments \cite{zhangCavityMagnomechanics2016b, pottsDynamicalBackactionMagnomechanics2021a} have involved magnetic spheres which yield mechanical modes of MHz frequencies, due to their relatively large diameters on the order of $100\,\mu \rm{m}$. This limits the parametric interaction strengths with a uniform magnetic Kittel mode (frequency on the order of GHz) to a few mHz. In contrast, resonant magnomechanical interaction yields much larger coupling strengths. For mode frequencies of $2\pi \times 10\,\rm{GHz}$ we expect an optimal coupling strength on the order of $g_{mb}=2\pi\times10\,\rm{MHz}$. If the mode volumes of the magnon ($V_m$) and phonon ($V_b$) do not match, the coupling scales as $\left(2B_1 +B_2\right)\rho^{-1/2}\sqrt{V_m/V_b}$ (if $V_m<V_b$) or as $\left(2B_1 +B_2\right)\rho^{-1/2}\sqrt{V_b/V_m}$ (if $V_m>V_b$), where $B_1=3.48\,\rm{J}\cdot \rm{m}^{-3}$ and $B_2=6.4\,\rm{J}\cdot \rm{m}^{-3}$ are the magnetoelastic constants for YIG \cite{gurevichMagnetizationOscillationsWaves2020} and $\rho=5110\,\rm{kg \cdot m^{-3}}$ is the mass density \cite{stancilSpinWavesTheory2010a}. Detailed derivations of all coupling terms can be found in the Appendix. \par
Magnons and microwave photons couple via magnetic dipole interactions. Experimentally, the coupling can be achieved by loading the magnetic element in a microwave cavity, and coupling strengths on the order of the modes frequencies have been achieved \cite{hueblHighCooperativityCoupled2013b, zhangBroadbandNonreciprocityEnabled2020a}. For perfect mode overlap, the coupling only scales with the material specific spin density. We estimate an optimal coupling strength of $1.85\,\rm{GHz}$ for mode frequencies of $10\,\rm{GHz}$. Nevertheless, such an optimal coupling requires perfect mode matching, which can be challenging for small magnon mode volumes. The difference between the magnon ($V_m$) and microwave mode ($V_c$) mode volumes yields a smaller coupling, which scales via $\sqrt{\rho_{\rm{S,YIG}}V_m/V_c}$ (if $V_c>V_m$), with $\rho_{\rm{S, YIG}}=4.22\times10^{27}\rm{m}^{-3}$ the spin density of YIG \cite{stancilSpinWavesTheory2010a}. In what follows, we assume the inductive coupling value of $g_{mc}=2\pi\times180\,\rm{MHz}$, corresponding to a 1 $\%$ magnon to microwave volume ratio (c.f. Appendix 3), achieved for example in \cite{pottsStrongMagnonPhoton2020}. In section \ref{subsec:smallinductive} we will further discuss the case of $g_{mc}=2\pi\times18\,\rm{MHz}$, corresponding to a volume ratio of only $0.01\%$, which should be more easily achievable in experiment. \par
The interaction between phonons and optical photons arises from photoelastic effects. A mechanical mode dynamically deforms an optical cavity, which affects the permittivity inside the material, therefore altering its resonance frequency. Maximizing the achievable coupling value for perfect mode overlap also requires the mode volumes, $V_a$ (photon) and $V_b$ (phonon), to be small, since the coupling scales with $n^4\rho^{-1/2}/\epsilon_r\sqrt{V}$ (for $V_a=V_b=V$), with $n=2.2$ and $\epsilon_r=5$ the refractive index and relative permittivity for YIG (both at $1.2\mu \text{m}$), respectively \cite{stancilSpinWavesTheory2010a}. 
If the mode volumes do not match, the coupling scales either with $n^4 \rho^{-1/2}/\epsilon_r\sqrt{V_b}$ (if $V_a<V_b$) or $n^4\rho^{-1/2}\sqrt{ V_b}/\epsilon_r V_a$ (if $V_a>V_b$). For YIG, we estimate an optimal single photon optomechanical coupling $g_{ab}=2 \pi \times 0.2\,\rm{MHz}$ for a mode volume of $1\,\mu {\rm{m}}^3$, which is $~80\%$ smaller than for state-of-the-art optomechanical crystals which are made of Silicon \cite{eichenfieldOptomechanicalCrystals2009a, chanOptimizedOptomechanicalCrystal2012b}. The main reason for that is the smaller refractive index of YIG compared to Silicon. Other material specific constants, such as the permittivity and mass density, are better in YIG but cannot compensate the quartic dependence on the refractive index (compare Eq. \eqref{eq:optomech_coup_est}). We are ignoring a moving-boundary contribution, since its contribution is smaller than the photoelastic one and difficult to estimate without considering a specific geometry. \par
The optical cavity is connected to a coherent drive, leading to the total Hamiltonian 
\begin{equation}
    \hat{H}_2=\hat{H}_1+\hat{H}_{\rm{drive}},
\end{equation}
with an optical drive term $\hat{H}_{\rm{drive}}=i\alpha\left(\hat{a}e^{i\omega_Dt}-\hat{a}^{\dagger}e^{-i\omega_Dt}\right)$. We perform an unitary transformation to the rotating frame of the optical driving field via $\hat{U}\hat{H}\hat{U}^{\dagger} -i \hbar \hat{U} \partial \hat{U} / \partial t$ with $\hat{U}=\exp{\left[i\omega_D \hat{a}^{\dagger}\hat{a} t\right]}$ and linearize the Hamiltonian by considering fluctuations on top of the coherent steady-state induced by the drive. For small coupling rates, only the optical field exhibits a coherent part, such that $\hat{a}=\alpha+\delta \hat{a}$ with $\alpha = \sqrt{\bar{n}_a}$, where $\bar{n}_a$ is the mean photon number in the optical cavity. The mean photon number $\bar{n}_a$ can be estimated by \cite{aspelmeyerCavityOptomechanics2014b}
\begin{equation}
    \bar{n}_a=\frac{\kappa_{a,\rm{out}}P}{\hbar \omega_a \left[\Delta_a ^2 + \left (\gamma_a + \kappa_{a,\rm{out}} \right)^2/4 \right]},
    \label{eq:meanphotonnumber}
\end{equation}
where $P$ is the drive power, $\gamma_a$ the internal optical mode linewidth, $\kappa_{a,out}$ the dissipation rate to the external port and $\Delta_a = \omega_D - \omega_a$ the detuning from the drive frequency. The confinement of photons yields a coupling enhancement of the parametric processes which scales with the square root of the mean photon number, specifically, the cavity enhanced coupling is given by $G_{ab}= \sqrt{\bar{n}_a} g_{ab}$. Next we apply a rotating wave approximation, neglecting all terms of the form  $\hat{m}^{\left(\dagger\right)}\hat{b}^{\left(\dagger\right)}$ and $\hat{m}^{\left(\dagger\right)}\hat{c}^{\left(\dagger\right)}$. As long as the coupling rates remain smaller than about $10 \%$ of the mode frequencies, so $G_{ab},g_{mb},g_{mc}< 0.1 \left(\Delta_a, \omega_m, \omega_b, \omega_c\right)$, they can be safely neglected \cite{friskkockumUltrastrongCouplingLight2019,forn-diazUltrastrongCouplingRegimes2019}. For a red-detuned drive $\Delta_a \approx -\omega_{b}$, the terms $\delta \hat{a}^{(\dagger)} \hat{b}^{(\dagger)}$ can also be neglected under the rotating wave approximation. This leads to the Hamiltonian used in the following (from now on we adopt the notation $\delta \hat{a} \rightarrow \hat{a}$)
\begin{eqnarray}
   \hat{H}_2/\hbar =&&-\Delta_a \hat{a}^{\dagger}\hat{a} + \omega_m \hat{m}^{\dagger}\hat{m} + \omega_b \hat{b}^{\dagger}\hat{b} + \omega_c \hat{c}^{\dagger}\hat{c}\nonumber\\
   && + G_{ab} \left( \hat{a}^{\dagger} \hat{b} + \hat{a} \hat{b}^{\dagger} \right)\nonumber
    + g_{mb} \left( \hat{m}^{\dagger} \hat{b} + \hat{m} \hat{b}^{\dagger} \right)\\ 
    &&+ g_{mc} \left( \hat{m}^{\dagger} \hat{c} + \hat{m} \hat{c}^{\dagger} \right).
    \label{eq:model_Hamiltonian}
\end{eqnarray}
\begin{table}
\caption{\label{tab:values}Table of used values for mode frequencies, linewidths/dissipations and coupling strengths.$g_{mb}$ and $g_{ab}$ assume a perfect mode overlap at a phonon, photon and magnon mode volume of $1\,\mu {\rm{m}}^3$. The estimate for $g_{mc}$ assumes $1\%$ mode overlap between the magnon and the microwave field, so the microwave mode volume is $100\,\mu {\rm{m}}^3$.}
\begin{ruledtabular}
\begin{tabular}{ccc}
 Quantity  &Symbol &Value \\ \hline
  Optical photon freq.                & $\nicefrac{\omega_a}{2\pi}$           & $200\,\rm{THz}$  \\ 
 
 Optical detuning (red)               & $\nicefrac{\Delta_a}{2\pi}$           & $-10\,\rm{GHz}$  \\
 
 Phonon freq.                        & $\nicefrac{\omega_b}{2\pi}$           & $10\,\rm{GHz}$  \\
 
 Magnon freq.                       & $\nicefrac{\omega_m}{2\pi}$           & $10\,\rm{GHz}$  \\

 Internal optical linewidth         & $\nicefrac{\gamma_a}{2\pi}$           & $0.1\,\rm{GHz}$  \\
 
 External optical dissipation       & $\nicefrac{\kappa_{a,\rm{out}}}{2\pi}$     & $1.5\,\rm{GHz}$  \\
 
 Phonon linewidth                   & $\nicefrac{\gamma_b}{2\pi}$           & $1.0\,\rm{kHz}$  \\
 
 Magnon linewidth                   & $\nicefrac{\gamma_m}{2\pi}$           & $1.0\,\rm{MHz}$  \\
 
 Internal mw cavity loss rate      & $\nicefrac{\gamma_c}{2\pi}$            & $3.0\,\rm{MHz}$  \\
 
 External mw cavity coupling rate     & $\nicefrac{\kappa_{c,\rm{out}}}{2\pi}$     & $100\,\rm{MHz}$  \\
 
 Magnomechanical coupling           & $\nicefrac{g_{mb}}{2\pi}$             & $10.0\,\rm{MHz}$  \\
 
 Magnon-Microwave photon coupling          & $\nicefrac{g_{mc}}{2\pi}$      & $180\,\rm{MHz}$  \\

 Optomechanical coupling            & $\nicefrac{g_{ab}}{2\pi}$             & $0.2\,\rm{MHz}$ \\
 
 Mode volume (photon, phonon, magnon)           & $V_{a/b/m}$             & $1\,\mu {\rm{m}}^3$\\
 
 Mode volume (microwave)           & $V_{c}$             & $100\,\mu {\rm{m}}^3$\\
\end{tabular}
\end{ruledtabular}
\end{table}
\section{Conversion setup \label{section:protocol}}
\begin{figure}
    \centering
    \includegraphics[width=0.48\textwidth]{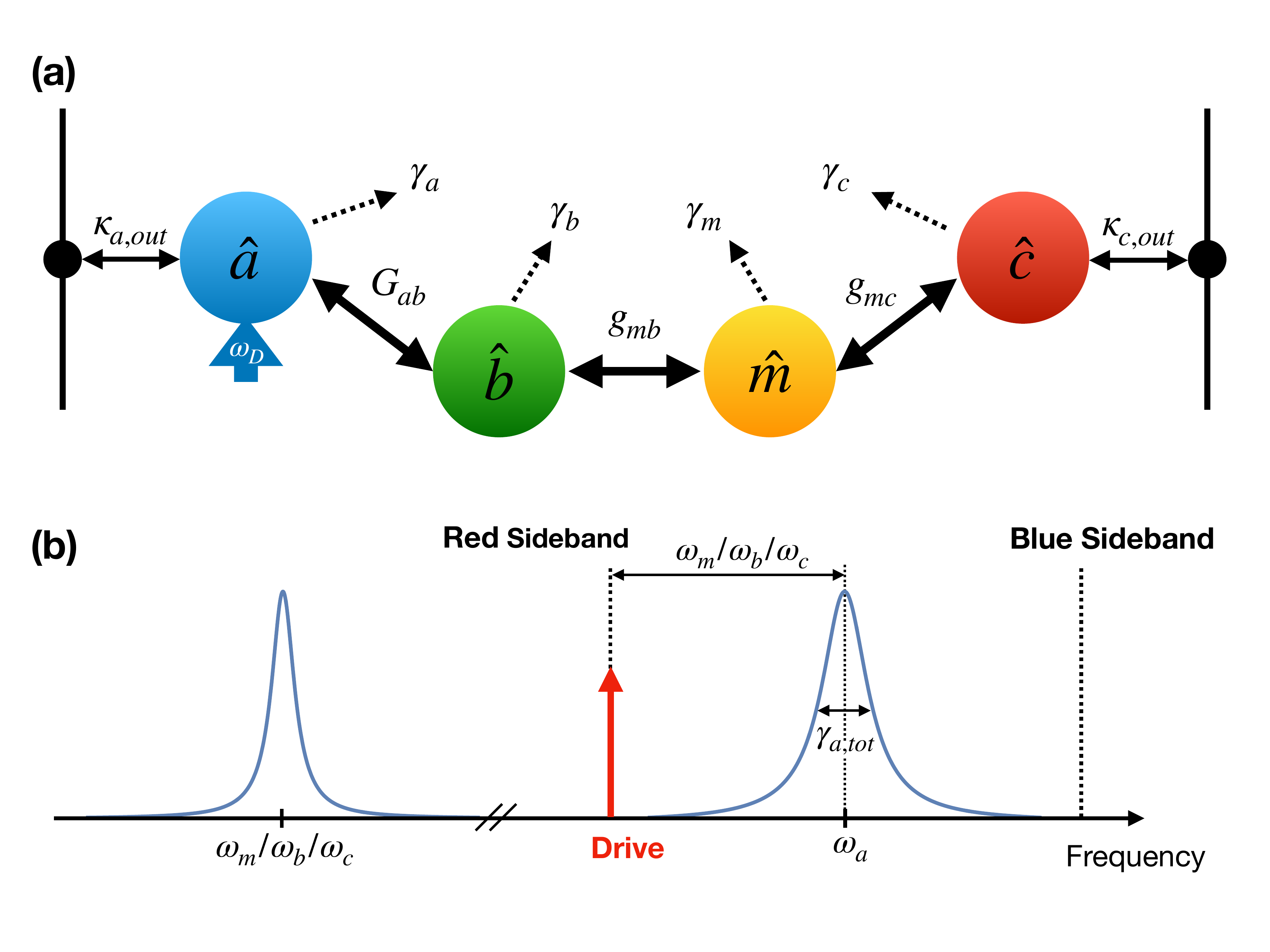}
    \caption{(a) Schematic and notation of the quadri-partite coherent interconversion protocol showing the interdependence of the four different waveforms. The chain arises from selecting the dominant coupling mechanisms in Fig. \ref{fig:model_interactionscheme}. Here, $\kappa_{a/c,\rm{out}}$ denotes coupling to the optical/microwave port and $\gamma_{a/b/m/c}$ are the intrinsic linewidth of each waveform. We consider an optical pump at frequency $\omega_D$ and therefore the optomechanical coupling is enhanced and denoted by $G_{ab}$. (b) Pumping scheme for driving of the red sideband of the optical cavity. Here $\omega_a, \omega_b, \omega_m$ and $\omega_c$ correspond to the optical photon, phonon, magnon and microwave photon mode, respectively.}
    \label{fig:protocol_conversionscheme}
\end{figure}

Based on the Hamiltonian of Eq. (\ref{eq:model_Hamiltonian}) we consider a conversion scheme as depicted in Fig. \ref{fig:protocol_conversionscheme} a). The quantum Langevin equations (QLE's) for the bosonic annihilation operators are obtained via the Heisenberg equation of motion $\hat{A}\left(t\right)=-i[\hat{A},\hat{H}]/\hbar$ (for an arbitrary operator $\hat{A}$), as
\begin{subequations}
\begin{eqnarray}
    \dot{\hat{a}}\left(t\right)=
    &&-i \left(-\Delta_a -i \frac{\gamma_{a, \rm{tot}}}{2}\right)\hat{a}\left(t\right) -i G_{ab}\hat{b}\left(t\right)-\nonumber\\
    &&\sqrt{\gamma_{a}}\hat{a}_{\rm{in,int}}\left(t\right)-\sqrt{\kappa_{a,\rm{out}}}\hat{a}_{\rm{in,ext}}\left(t\right),\label{eq:protocol_langevin_a}\\
    \dot{\hat{b}}\left(t\right)=
    && -i \left(\omega_b -i \frac{\gamma_b}{2}\right)\hat{b}\left(t\right) -i G_{ab}\hat{a}\left(t\right)\nonumber\\
    &&-i g_{mb}\hat{m}\left(t\right)-\sqrt{\gamma_{b}}\hat{b}_{\rm{in, int}}\left(t\right),\\
    \dot{\hat{m}}\left(t\right)=
    && -i \left(\omega_m -i \frac{\gamma_m}{2}\right)\hat{m}\left(t\right) -i g_{mb}\hat{b}\left(t\right)\nonumber\\ 
    &&-i g_{mc}\hat{c}\left(t\right)-\sqrt{\gamma_{m}}\hat{m}_{\rm{in, int}}\left(t\right),\\
    \dot{\hat{c}}\left(t\right)=
    &&-i \left(\omega_c -i \frac{\gamma_{c, \rm{tot}}}{2}\right)\hat{c}\left(t\right) -i g_{mc}\hat{m}\left(t\right)\nonumber\\
    && -\sqrt{\gamma_{c}}\hat{c}_{\rm{in,int}}\left(t\right)\left(t\right)-\sqrt{\kappa_{c,\rm{out}}}\hat{c}_{\rm{in,ext}}\left(t\right),
    \label{eq:protocol_langevin_d}
\end{eqnarray}
\end{subequations}
where we have included the internal noise operators $\hat{a}_{\rm{in,int}}$, $\hat{b}_{\rm{in,int}}$, $\hat{m}_{\rm{in,int}}$, $\hat{c}_{\rm{in,int}}$ to describe the open dynamics of the system, and $\hat{a}_{\rm{in,ext}}$, $\hat{c}_{\rm{in,ext}}$ for the input fields from the external ports. The description of the mechanical mode in terms of the annihilation operator $\hat{b}$ instead of the quadrature $\hat{x}$ is justified since the setup operates in the regime $\omega_b \gg \gamma_b$ \cite{aspelmeyerCavityOptomechanics2014b} (compare Tab. \ref{tab:values}). The rates $\gamma_{a/b/m/c}$ define the internal linewidths of the modes. Furthermore, $\gamma_{a/c, \rm{tot}}=\gamma_{a/c}+\kappa_{a/c,\rm{out}}$ gives the total linewidth of the optical/microwave mode which includes the coupling rates $\kappa_{a/c,\rm{out}}$ to the two external ports. 
All noise operators $\hat{\beta}=\hat{a}_{\rm{in,int}}, \hat{a}_{\rm{in,ext}}, \hat{b}_{\rm{in,int}}, \hat{m}_{\rm{in,int}}, \hat{c}_{\rm{in,int}}, \hat{c}_{\rm{in,ext}}$ describe thermal noise in each mode, given by
\begin{eqnarray}
    \langle \hat{\beta}\left(t\right)\hat{\beta}^{\dagger}\left(t^{\prime}\right)\rangle && =\left(n_{th,\beta}+1\right)\delta\left(t-t^{\prime}\right),\nonumber\\
    \langle \hat{\beta}^{\dagger}\left(t\right)\hat{\beta}\left(t^{\prime}\right)\rangle && =n_{th,\beta}\delta\left(t-t^{\prime}\right),
\end{eqnarray}
where $n_{th,\beta}=\left[{\rm{exp}}\left(\hbar \omega_{\beta} /k_{\rm{B}}T\right)-1\right]^{-1}$ is the Bose-Einstein distribution determining the thermal occupancy of the respective baths, with $k_{\rm{B}}$ the Boltzmann constant and $T$ the temperature. For the optical photons the thermal occupancy is $n_{th,a}\approx 0$ even at room temperature due to the high frequency.
In the next step, we define the $4\times4$ matrix $A$ and the $4\times6$ matrix $B$ to bring the QLE's in the form
\begin{equation}
    \dot{\textbf{v}}\left(t\right)=A\textbf{v}\left(t\right)+B\textbf{v}_{in}\left(t\right),
    \label{eq:protocol_matrixform}
\end{equation}
with $\textbf{v}\left(t\right)=\left(\hat{a},\hat{b},\hat{m},\hat{c}\right)^T$ and $\textbf{v}_{\rm{in}}\left(t\right)=\left(\hat{a}_{\rm{in,ext}},\hat{a}_{\rm{in,int}},\hat{b}_{\rm{in}},\hat{m}_{\rm{in}},\hat{c}_{\rm{in,ext}},\hat{c}_{\rm{in,int}}\right)^T$, where we distinguish between internal and external dissipation paths for optical photons and microwaves. \par
In order to obtain the conversion spectrum of our coupled system, we compute the scattering matrix $S\left(\omega\right)$ which relates input and output in frequency space via $\textbf{v}\left(\omega\right)_{\rm{out}}=S\left(\omega\right)\textbf{v}_{\rm{in}}\left(\omega\right)$. For that, we Fourier transform the QLE's (Eqs. (\ref{eq:protocol_langevin_a}) to (\ref{eq:protocol_langevin_d})) using the convention
\begin{subequations}
\begin{eqnarray}
   && \int_{-\infty}^{\infty} dt\, e^{i\omega t} \hat{O}\left(t\right)= \, \hat{O}\left(\omega\right),\\
    && \int_{-\infty}^{\infty} dt\, e^{i\omega t} \dot{\hat{O}}\left(t\right)= -i\omega \hat{O}\left(\omega\right),
\end{eqnarray}
\end{subequations}
and obtain
\begin{subequations}
\begin{eqnarray}
    \hat{a}=&&-i G_{ab}\chi_a [\omega] \hat{b}-\sqrt{\gamma_{a}}\chi_a[\omega]\hat{a}_{\rm{in, int}}\nonumber\\&&
    -\sqrt{\kappa_{out,a}}\chi_a[\omega]\hat{a}_{\rm{in, out}},\label{eq:protocol_langevin_frq_a}\\
    \hat{b}=&&-i G_{ab}\chi_b [\omega] \hat{a}-ig_{mb}\chi_b[\omega]\hat{m}\nonumber\\&&-\sqrt{\gamma_b}\chi_b[\omega]\hat{b}_{\rm{in}},\label{eq:protocol_langevin_frq_b}\\
    \hat{m}=&&-i g_{mb}\chi_m [\omega] \hat{b}-ig_{mc}\chi_m[\omega]\hat{c}\nonumber\\&&-\sqrt{\gamma_m}\chi_m[\omega]\hat{m}_{\rm{in}},\label{eq:protocol_langevin_frq_c}\\
    \hat{c}=&&-i g_{mc}\chi_c [\omega] \hat{m}-\sqrt{\gamma_{c}}\chi_c[\omega]\hat{c}_{\rm{in, int}}\nonumber\\&&
    -\sqrt{\kappa_{out,c}}\chi_c[\omega]\hat{c}_{\rm{in, ext}},
    \label{eq:protocol_langevin_frq_d}
\end{eqnarray}
\end{subequations}
where the susceptibilities have the form e.g. \\ $\chi_c[\omega]=\left(-i\left(\omega - \omega_c\right)+\left(\gamma_c+\kappa_{c,out}\right)/2\right)^{-1}$. We obtain the scattering matrix as a function of frequency $\omega$ referenced to a red sideband tone via
\begin{equation}
    S\left(\omega\right)=B^T \left[-i\omega I_4 - A\right]^{-1}B - I_6,
    \label{eq:scattering_matrix}
\end{equation}
where $I_N$ is the $N\times N$ identity matrix. The conversion efficiency from optics to microwave is then obtained as $\xi_{ac}\left(\omega\right)= \lvert S_{1,5}\left(\omega\right)\rvert^2$ and explicitly reads
\begin{widetext}
\begin{equation}
 \xi_{ac}\left(\omega\right)= {\Big \lvert} \frac{\chi_{a}[\omega]\chi_{b}[\omega]\chi_{m}[\omega]\chi_{c}[\omega] G_{ab} g_{mc} g_{mb} \sqrt{\kappa_{\rm{out},a}\kappa_{\rm{out},c}}}
    {(1 + g_{mc}^2\chi_{m}[\omega]\chi_{c}[\omega])(1+ G_{ab}^2\chi_{a}[\omega]\chi_{b}[\omega]) +g_{mb}^2 \chi_{m}[\omega]\chi_{b}[\omega]} {\Big\rvert}^2 .\;
    \label{eq:efficiency_magnomech_omega}
\end{equation}
\end{widetext}
\par
In the case that all modes are on resonance and the cavity is driven with a red detuning (see Fig. \ref{fig:protocol_conversionscheme} (b)), $\omega=\omega_m=\omega_b=\omega_c=-\Delta_a$, we can express Eq. (\ref{eq:efficiency_magnomech_omega}) solely in terms of the cooperativities $C_{ij}=4g_{ij}^2 /\gamma_{i,\rm{tot}} \gamma_{j,\rm{tot}}$
\begin{equation}
    \xi_{ac}=\eta_a \eta_c \frac{4C_{ab}C_{mb}C_{mc}}{\left[(1+C_{ab})(1+C_{mc})+C_{mb}\right]^2},
    \label{eq:protocol_magnomech_convo}
\end{equation}
where $\eta_{a/c}=\nicefrac{\kappa_{a/c,\rm{out}}}{\gamma_{a/c, \rm{tot}} }$ are the extraction factors defined as the ratio between port dissipation and total dissipation. A cooperativity value of $C_{ij}\gg 1$ is a necessary criterion for an efficient energy transfer \cite{barzanjehReversibleOpticaltoMicrowaveQuantum2012, wangUsingInterferenceHigh2012}. \par
\begin{figure}
    \centering
    \includegraphics[width=0.48\textwidth]{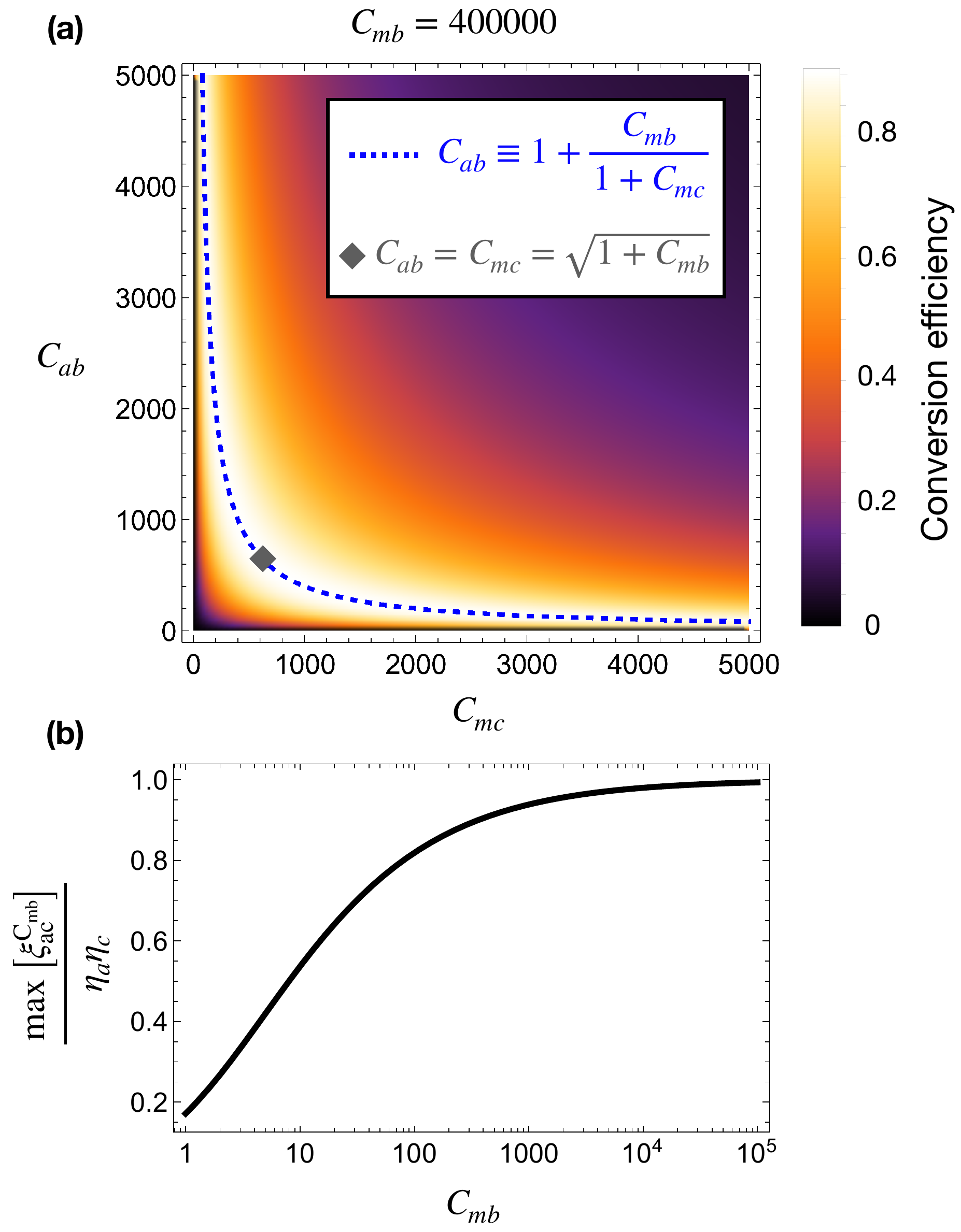}
    \caption{(a) Conversion efficiency $\xi$ at full degeneracy between the different microwave waveforms ($\omega=\omega_m=\omega_b=\omega_c=-\Delta_a$). The expression is given by Eq. (\ref{eq:protocol_magnomech_convo}) as a function of the cooperativity between magnon and microwave photon ($C_{mc}$) and optical photon and phonon ($C_{ab}$). The dashed line in the lower plot denotes the maximum conversion efficiency. The grey dot marks the global maximum. The magnon-phonon cooperativity is fixed at $C_{mb}=4\times10^5$, corresponding to the estimates of Table \ref{tab:values}. (b) Maximum conversion efficiency given in Eq. \eqref{eq:max_coop_cmb} as a function of $C_{mb}$. $C_{ab}$ and $C_{mc}$ fulfil Eq. \eqref{eq:extremal_coop_opt}.}
    \label{fig:appendix_coop}
\end{figure}

The single-particle couplings and intrinsic decay rates are fixed by the design of the system and by the material. Therefore, we consider a setup in which the free variables are the photon-phonon cooperativity $C_{ab}$, which can be tuned via the power of the external optical drive, and the magnon-microwave cooperativity $C_{mc}$, which can be changed by adjusting the coupling of the microwave cavity to the external port. By setting the partial derivatives of Eq. \eqref{eq:protocol_magnomech_convo} with respect to $C_{mc}$ and $C_{ab}$ to zero while fixing $C_{mb}$, we obtain a constraint for the cooperativities
\begin{equation}
    C_{ab}\equiv 1+\frac{C_{mb}}{C_{mc}+1},
    \label{eq:perfectEffCondition}
\end{equation}
which in turns gives the maximum efficiency at
\begin{equation}
    C_{ab}^{\rm{max}}=C_{mc}^{\rm{max}}=\sqrt{1+C_{mb}}.
    \label{eq:extremal_coop_opt}
\end{equation}
The optimal value for the conversion efficiency in terms of the fixed cooperativity $C_{mb}$ is then given by
\begin{equation}
    {\rm{max}}\left[\xi_{ac}^{C_{mb}}\right]=\eta_a\eta_c\frac{(1 + C_{mb})  C_{mb}}{\left(1+C_{mb} + \sqrt{1 + C_{mb}}\right)^2},
    \label{eq:max_coop_cmb}
\end{equation}
which approaches $\eta_a\eta_c$ as $C_{mb}$ grows. In Fig. \ref{fig:appendix_coop}(a), we show the conversion efficiency as a function of the cooperativities $C_{ab}$ and $C_{mc}$ at $C_{mb}=4\times 10^5$, corresponding to our estimates shown in Table \ref{tab:values}. The indicated global maximum is obtained from Eq. \eqref{eq:max_coop_cmb}, which for our adopted parameters yields a maximum conversion efficiency of $\sim 0.997 \times \eta_a \eta_c$. The optimal cooperativities obtained with Eq. \eqref{eq:extremal_coop_opt} determine the required microwave port dissipation  $\kappa_{c}^{\rm{max}}\approx2\pi\times202\,\rm{MHz}$ and an optical pump power $P_{\rm{max}}\approx 5.6\times 10^{-5}\,$ W, corresponding to $\approx6300$ circulating photons (see Eq. \eqref{eq:meanphotonnumber}) at a total optical linewidth of $\gamma_{a,\rm{tot}}=2\pi\times1.6\,\text{GHz}$. However, as long as Eq. \eqref{eq:perfectEffCondition} is fulfilled, high efficiencies are still possible even if Eq. \eqref{eq:extremal_coop_opt} is not satisfied. \par
It is important to note that the discussion exclusively in terms of cooperativities only holds true in the case where all mode frequencies (and the optical detuning) are on resonance. In a non-resonant case, the optimal conversion efficiency has to be found by optimizing Eq. (\ref{eq:efficiency_magnomech_omega}) with respect to the relevant parameters. This can be achieved by following the same procedure outlined above, but potentially requires numerical solutions for the obtained set of equations. \par  
When we leave out the phonon mode for the conversion scheme, transduction takes place only through the magnon mode, via magneto-optical effects and magnon-microwave coupling. This setup has been used in recent experiments \cite{hisatomiBidirectionalConversionMicrowave2016a, zhuWaveguideCavityOptomagnonics2020a}. In such a conversion scheme with only one mediating element, the efficiency in terms of the cooperativities reads \cite{ruedaEfficientMicrowaveOptical2016a}
\begin{equation}
    \eta_1 \eta_2 \frac{4C_{am} C_{mc}}{\left(1+C_{am}+C_{mc}\right)^2},
    \label{eq:protocol_magnonics_convo}
\end{equation}
where $C_{am}$ it the magnon-optical photon cooperativity. Compared with the efficiency via optomagnomechanics, given in Eq. (\ref{eq:protocol_magnomech_convo}), the efficiency for a single mediator requires matching cooperativities for maximum conversion. Since the optomagnonic coupling achieved experimentally is very low \cite{haighTripleResonantBrillouinLight2016a, zhangOptomagnonicWhisperingGallery2016} compared to magnon-microwave coupling, it is difficult to approach a good matching, even for very high pump powers. This limits the efficiencies obtained so far to values between $10^{-11}$ \cite{hisatomiBidirectionalConversionMicrowave2016a} and $10^{-7}$ \cite{zhuWaveguideCavityOptomagnonics2020a}. \par
The inclusion of the optomagnonic interaction in the magnomechanical-based setup considered here, adds a parallel conversion channel. When the optomagnonic and the optomechanical coupling strengths are comparable, destructive interference generates a Fano-shaped resonance in the conversion spectrum. In this case, the conversion efficiency vanishes at a specific frequency that depends on the relative strength of the optomagnonic and optomechanical couplings. 
\subsection{Hybridization and resonances \label{section:chainhybrid}}
\begin{figure*}
    \centering
    \includegraphics[width=1.0\textwidth]{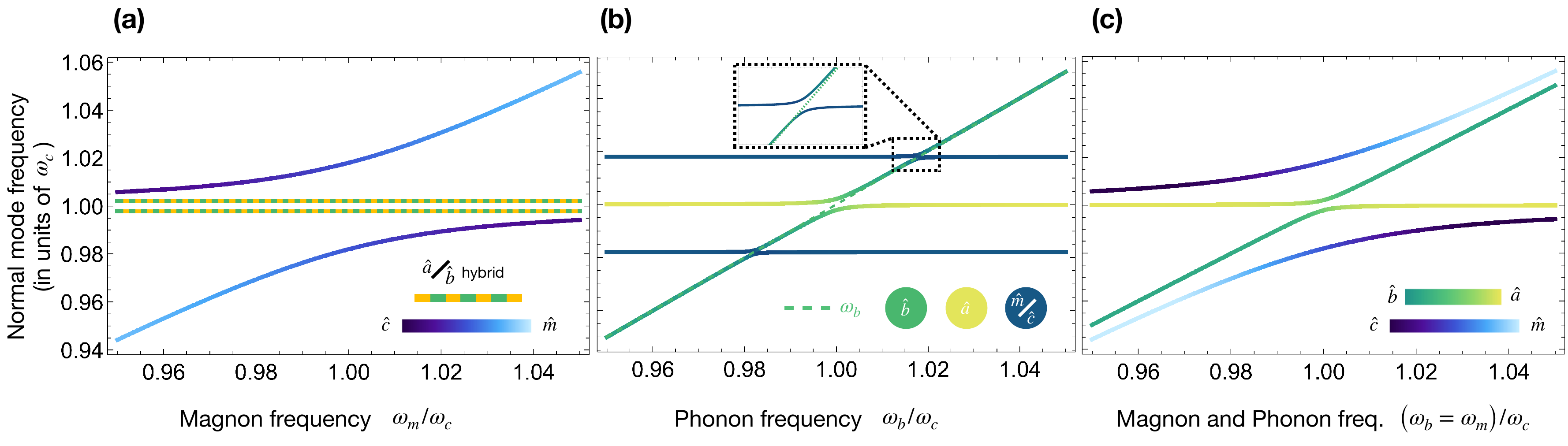}
    \caption{Plot of the hybridization ratio of the four waveforms as a function of their detuning. The avoided crossing region is shown in the rotating frame of the driven optical cavity, stemming from the interaction chain shown in Fig. \ref{fig:protocol_conversionscheme}, as a function of (a) the magnon frequency $\omega_m$, (b) the phonon frequency $\omega_b$ and (c) both set equal and simultaneously tuned. The color code corresponds to the weights of the different bare modes, and shows how the hybrid mode is formed as a combination of the bare modes. We normalized all frequencies by the fixed microwave frequency and set $\omega_c=-\Delta_a$. Parameters: optical pump power $P=10^{-4}$ W, corresponding to a photon-phonon cooperativity $C_{ab}=1124$. Magnon-phonon cooperativity $C_{mb}=4\times 10^5$, and magnon-microwave cooperativity $C_{mc}=1258$, corresponding to the estimates presented in Tab. \ref{tab:values}.}
    \label{fig:protocol_chain_hybridization}
\end{figure*}
Coupled systems, where the coupling strength between the interacting modes is larger than their linewidths, are considered in the strong coupling regime. Based on our coupling estimates and anticipated linewidths (see Tab. \ref{tab:values}), all considered modes are strongly coupled, except phonons and optical photons. Here, the coupling regime depends on the optical pump power. Strongly coupled systems undergo normal mode splitting when they are close to resonance. In particular, for a system with only two coupled modes $i$ and $j$ at $\omega_i=\omega_j=\omega$ the shifted eigenfrequencies are given by $\omega_{\pm}=\omega \pm g_{ij}$, where $g_{ij}$ is the coupling strength. If the total linewidth $\gamma_{j,\rm{tot}}$ of a coupled mode $j$ exceeds the coupling strength $g_{ij}$, the shifted hybrid peaks cannot be resolved. If, however, $4g_{ij}>\gamma_{i,\rm{tot}},\gamma_{j,\rm{tot}}$, the system is considered to be in the strong coupling regime, and the two shifted hybrid peaks can be clearly resolved.\par
In Fig. \ref{fig:protocol_chain_hybridization}(a) we show the normal mode spectrum of Eq. \eqref{eq:model_Hamiltonian} of the interaction chain shown in Fig. \ref{fig:protocol_conversionscheme} (a). At the resonance point ($\omega_c=\omega_b=\omega_m=-\Delta_a$) we obtain two sets of hybrid modes - one mode partly magnon and microwave and the other partly phonon and optical photon. Since the coupling between magnons and microwaves is the dominant coupling, the phonons can only hybridize with the driven optical cavity mode. Hybridization between magnons and phonons does not occur as long as magnon-microwave interaction dominates. However, if we tune the phonon frequency instead (Fig. \ref{fig:protocol_chain_hybridization}(b)), we can bring it in resonance with the magnon-microwave hybrid mode. The phonon then couples with the magnon-part and the spectrum is split by twice the magnomechanical coupling rate $g_{mb}$. Simultaneous tuning of magnon and phonon frequency in the case $\omega_b = \omega_m$, as shown in Fig. \ref{fig:protocol_chain_hybridization}(c), only shows magnon-phonon hybridization when both are tuned away from the microwave frequency.\par
\begin{figure}
    \centering
    \includegraphics[width=0.48\textwidth]{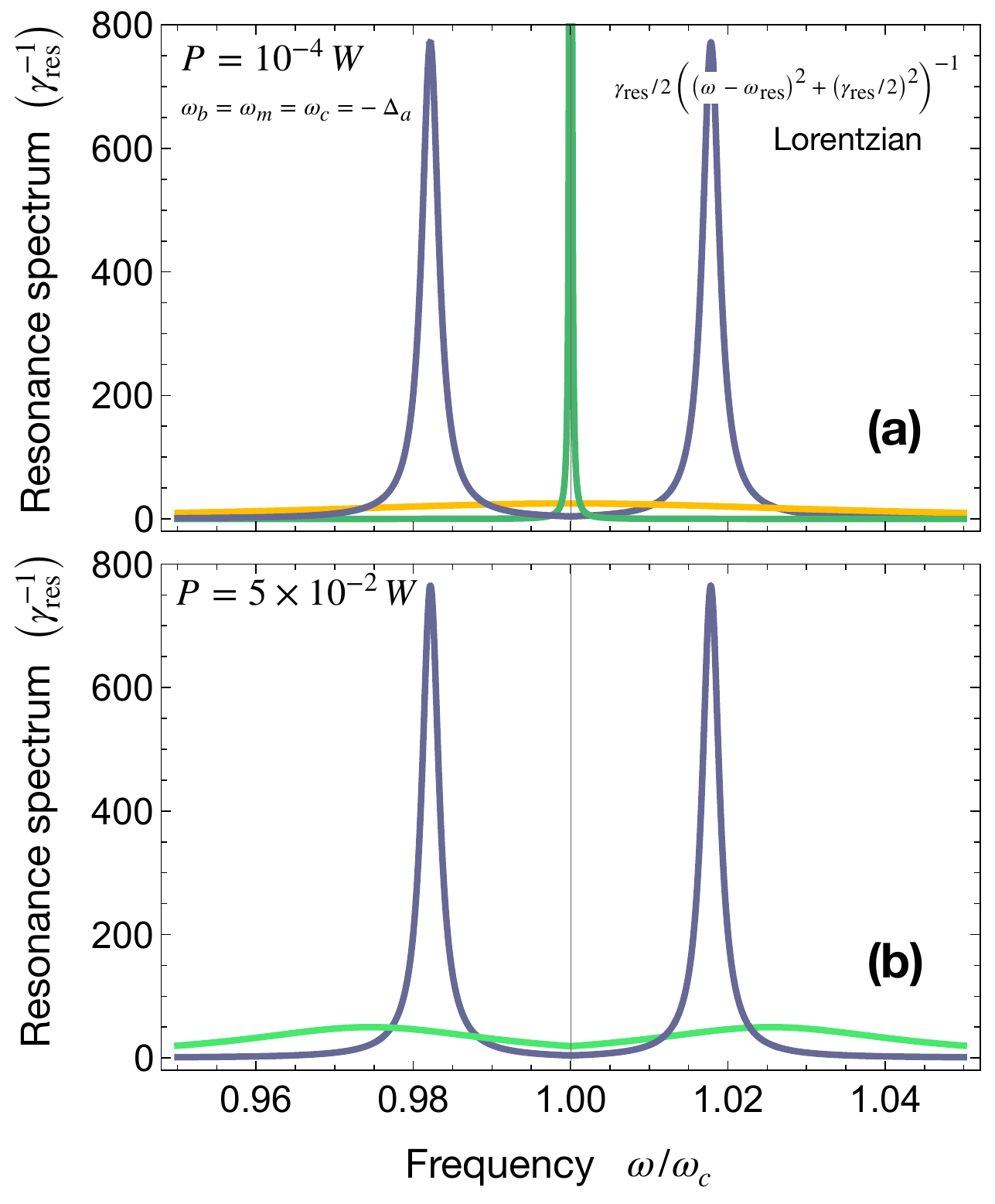}
    \caption{Frequency spectrum of the the resonances of the coupled system defined by Eqs. (\ref{eq:protocol_langevin_frq_a}) to (\ref{eq:protocol_langevin_frq_d}). The spectrum is obtained when all modes are on resonance including the optical detuning. The blue peaks correspond to the magnon-microwave hybrid modes. We assume full mode degeneracy ($\omega=\omega_m=\omega_b=\omega_c=-\Delta_a$) and, based on the estimates of Tab. \ref{tab:values}, we have fixed cooperativites at $C_{mb}=4\times10^{5}$ and $C_{mc}=1258$ for magnomechanics and magnon-microwave coupling, respectively. The spectrum is depicted for a pump power of (a) $10^{-4}$ ($C_{ab}=1124$) and (b) $5\times10^{-2}\rm{W}$ ($C_{ab}=562353$). In (a) the system is in the weak optomechanical coupling regime, hence no resolved shift of the phonon (green) and photon (yellow) mode. In (b) the system exhibits strong optomechanical coupling and therefore a normal mode splitting (light green) becomes evident.}
    \label{fig:protocol_hyridfreq_spectrum}
\end{figure}
\begin{figure}
    \centering
    \includegraphics[width=0.48\textwidth]{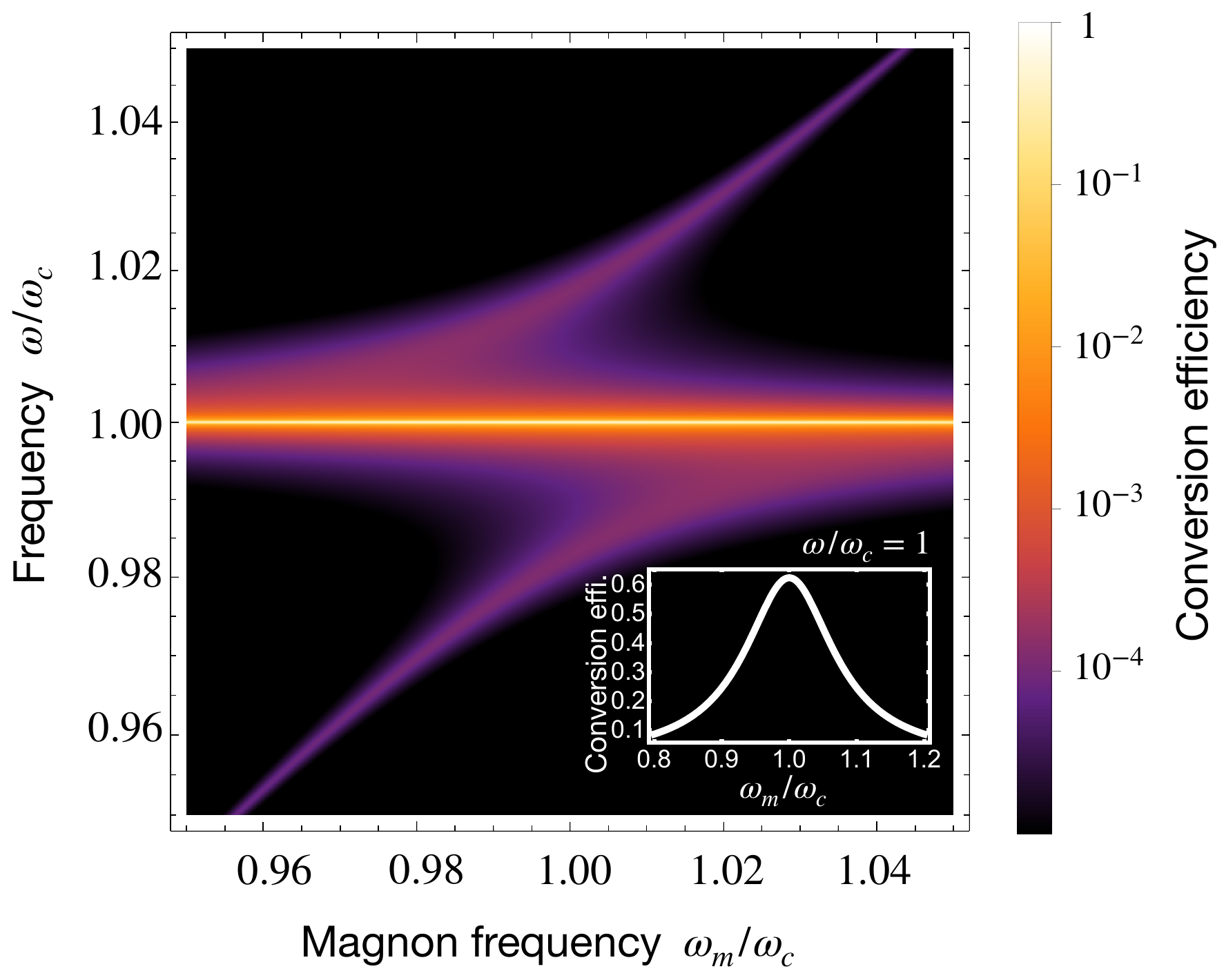}
    \caption{Conversion efficiency, Eq. (\ref{eq:efficiency_magnomech_omega}), as function of frequency $\omega$ and magnon frequency $\omega_{m}$, both in units of the fixed microwave frequency $\omega_{c}$. The phonon frequency $\omega_b$ and optical detuning are set to $-\omega_c$. The white inset shows the conversion efficiency at $\omega/\omega_{c}=1$ as a function of magnon frequency. The optical pump power was set to $P=10^{-4}\rm{W}$ corresponding to the cooperativity $C_{ab}=1124$. Other parameters fixed according with the estimates of Tab \ref{tab:values}, corresponding to $C_{mb}=4\times 10^{5}$ and $C_{mc}=1258$.}
    \label{fig:densityplt_effHybridization}
\end{figure}

To include the linewidths of the bare modes, we look at the resonances of the coupled system of Eqs. (\ref{eq:protocol_langevin_frq_a}) - (\ref{eq:protocol_langevin_frq_d}). The imaginary part of the eigenvalues corresponds to a resonance frequency $\omega_{\rm{res}}$, while the real part gives the width of the resonance peak $\gamma_{\rm{res}}$. The resonance is given by the Lorentzian $\nicefrac{\gamma_{\rm{res}}}{2}\left(\left(\omega-\omega_{\rm{res}}\right)^2+\left(\gamma_{\rm{res}}\right)^2\right)^{-1}$. In Fig. \ref{fig:protocol_hyridfreq_spectrum} we plot the resonance spectrum for an optical pump power of $10^{-4}\,$ W and $5\times10^{-2}\,$ W. This enhances the optomechanical coupling rate to about $21\,\text{MHz}$ and $474\,\text{MHz}$, respectively. The latter case corresponds to the strong optomechanical coupling regime, where $4G_{ab}>\gamma_{a,\rm{tot}}$ \cite{aspelmeyerCavityOptomechanics2014b}. The total optical linewidth is set to $\gamma_{a,\rm{tot}}=2\pi\times1.6\,\text{GHz}$ (corresponding to a quality factor of $Q=125000$). As expected, in the first case ($P=10^{-4}\,$ W) we observe no hybridization between photon and phonon, hence no shift of the green and yellow resonance peaks. Conversely, the resonance peaks of the magnon-microwave hybrid modes are clearly resolved and separated by twice the coupling rate of $g_{mc}=2\pi\times180\,\text{MHz}$. In the second case ($P=5\times10^{-2}\,$ W) the transition to the strong optomechanical coupling regime also leads to two frequency shifted peaks. The peaks in the resonance spectrum correspond to the peaks of the conversion efficiency. This is, for example, depicted in Fig. \ref{fig:densityplt_effHybridization}, which shows the conversion efficiency as a function of frequency $\omega$ and magnon frequency $\omega_m$. One can observe a resemblance to Fig. \ref{fig:protocol_chain_hybridization}(a). The largest conversion efficiency is given at the point of full resonance, which is highlighted in the inset. Optimization of the maximum conversion efficiency would be performed following the discussion in terms of the cooperativities presented before.
\subsection{Conversion bandwidth}
\begin{figure*}
    \centering
    \includegraphics[width=1.0\textwidth]{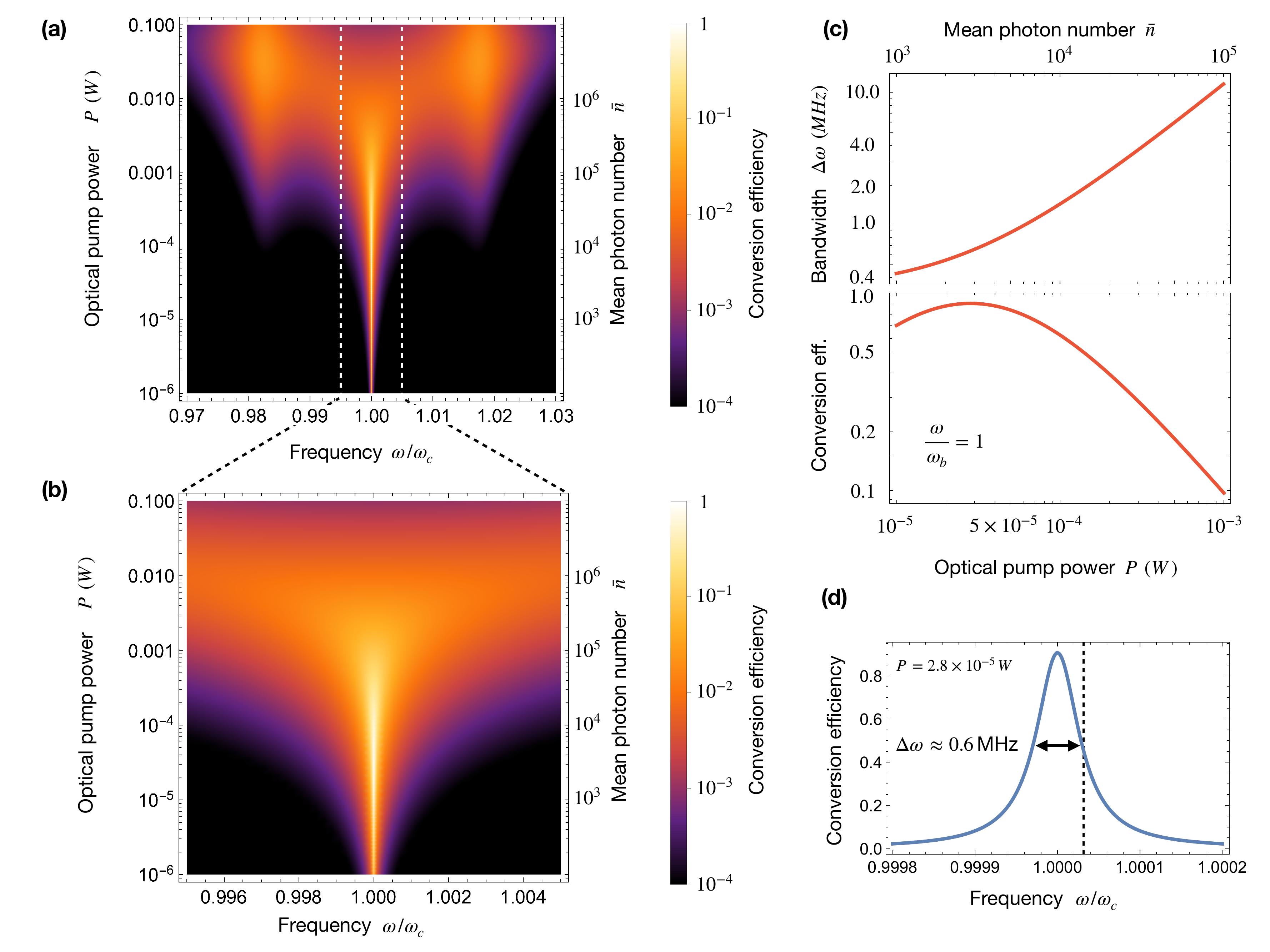}
    \caption{(a) Conversion efficiency as a function of frequency and optical pump power (corresponding mean photon number on right y-axis). The frequency axis is given in units of the microwave  mode frequency $\omega_c$ and the color code for the conversion efficiency is given on a logarithmic scale. We have set $\omega_b = \omega_m =\omega_c = -\Delta_a$. The lower plot (b) is a zoom-in of the upper plot. (c) Bandwidth and the respective conversion efficiency as a function of optical pump power. (d) Conversion efficiency as a function of frequency for the point of maximum efficiency at a pump power of $2.8\times10^{-5}\,\rm{W}$, corresponding to a cooperativity $C_{ab}=320$. The bandwidth at this point is marked with $0.6\,\rm{MHz}$. Other parameters are in correspondence with the estimates shown in Tab. \ref{tab:values} for which $C_{mb}=4\times10^5$ and $C_{mc}=1258$. }
    \label{fig:protocol_bandwidth_comp}
\end{figure*}
In order to characterize the performance of the proposed conversion system, we quantify the conversion bandwidth at frequencies with high conversion efficiency. In general, the conversion bandwidth is limited by the slowest process in the conversion chain. This is what limits e.g. state-of-the-art electromechanical and optomechanical systems whose bandwidths are defined by the small total linewidths of the vibrational modes \cite{ andrewsBidirectionalEfficientConversion2014a, higginbothamHarnessingElectroopticCorrelations2018a, vainsencherBidirectionalConversionMicrowave2016}. Conversely, modern broadband conversion setups suffer from small overall conversion efficiencies due to weak coupling \cite{zhuWaveguideCavityOptomagnonics2020a}. \par    
The conversion bandwidth can be calculated from the scattering matrix element given in Eq. (\ref{eq:efficiency_magnomech_omega}). The half width at full maximum (HWFM) $\Delta \omega_{\nicefrac12}$ of a peak in the transmission spectrum at a frequency $\omega_{\rm{max}}$ is defined by
\begin{equation}
    \xi\left(\Delta\omega_{\nicefrac12}\right)=\frac{\xi\left(\omega_\text{max}\right)}{2}.
\end{equation}
The conversion bandwidth is therefore defined as $\Delta\omega=2\Delta\omega_{\nicefrac12}$. Conversion spectra can be obtained using Eq. \eqref{eq:efficiency_magnomech_omega}. In Fig. \ref{fig:protocol_bandwidth_comp}(a) (Fig. \ref{fig:protocol_bandwidth_comp}(b) zoomed in) we show a plot of the conversion efficiency as a function of frequency and optical pump power (mean photon number). Based on the parameters in Tab. \ref{tab:values}, for small pump powers we notice high efficiency (close to unity) at the point of full resonance ($\omega_b = \omega_m = \omega_c = -\Delta_a $) with a bandwidth $\Delta\omega$ of up to $0.6\,\rm{MHz}$ at an optical pump power of $2.8\times10^{-5}\rm{W}$, which corresponds to the point of maximum conversion efficiency (see Fig. \ref{fig:protocol_bandwidth_comp}(d)). As the pump power increases, the resonance peak widens while the efficiency decreases. At pump powers $\sim10^{-3}\,\rm{W}$, conversion at the sidebands created from the strong magnon-microwave coupling are enhanced. This also marks the transition to the strong optomechanical coupling regime. In this region, we obtain a very large bandwidth (several hundred $\text{MHz}$) at the expense of a reduced conversion efficiency that does not drop below $10^{-4}$. It reaches $10^{-2}$ peak efficiency at the magnon-microwave sidebands that are limited in width by the microwave port coupling rate $\kappa_c$. Since $\kappa_c \gg \gamma_c$, the width is $\approx \nicefrac{\kappa_c}{2}=2\pi \times 50\,\rm{MHz}$. The bandwidth at $\omega_b = \omega_m = \omega_c = -\Delta_a $ can be increased up to about $10\,\text{MHz}$ before the efficiency drops off to $10^{-1}$ (see Fig. \ref{fig:protocol_bandwidth_comp}(c)). \par
Optimization of the conversion bandwidth requires improvement of the limiting coupling processes in the chain. In the weak optomechanical coupling regime, where the conversion efficiency is maximum, the bandwidth is limited by the smallest mode coupling rate, which is either $g_{mb}$ or $G_{ab}$, depending on the pump power. In the strong optomechanical regime, the bandwidth is increased by enhancing the coupling to the microwave port. Altering the magnon-microwave coupling rate only determines the spectral position of the sidebands, not their width. An enhancement of conversion bandwidth is achieved at the expense of reduced conversion efficiency. To optimize both, one has to take into account the interplay of the cooperativities between the coupled modes (Fig. \ref{fig:appendix_coop}) and find the optimal configuration. \par 
\subsection{Resolved vs. unresolved sideband regimes}
A large density of circulating photons in the optical resonator enhances the effective optomechancial coupling, but can be limited by the maximum amount of photons that the cavity supports. However, this limitation is partially lifted under optimization of the single photon-phonon coupling. Furthermore, Eq. (\ref{eq:meanphotonnumber}) implies that, aside from the pump power, the total optical linewidth has a crucial impact on the mean photon number for a given mode. In general, cavities with larger linewidths (low quality factors) require stronger pump powers for achieving a large density of circulating photons. \par
For the driven optical cavity we have to further distinguish between two regimes of the optical cavity with respect to the optical linewidth. In the resolved sideband regime, the total optical linewidth $\gamma_{a,\rm{tot}}=\gamma_a+\gamma_{a,\rm{out}}$ is smaller than the mode frequencies of magnons, microwave, and phonons: $\gamma_{a,\rm{tot}}<\omega_{b}$. In this case, the sidebands can be directly addressed by choosing the optical drive frequency (and therefore the optical detuning) to be $\pm\omega_b$. (compare Fig. \ref{fig:protocol_conversionscheme} (b)). If, however, $\gamma_{a,\rm{tot}}>\omega_{b}$ the sidebands cannot be resolved and therefore can not be individually addressed by a detuned drive. As a consequence, the precise value of the detuning has less impact on the obtained conversion efficiency. \par
\begin{figure}
    \centering
    \includegraphics[width=0.48\textwidth]{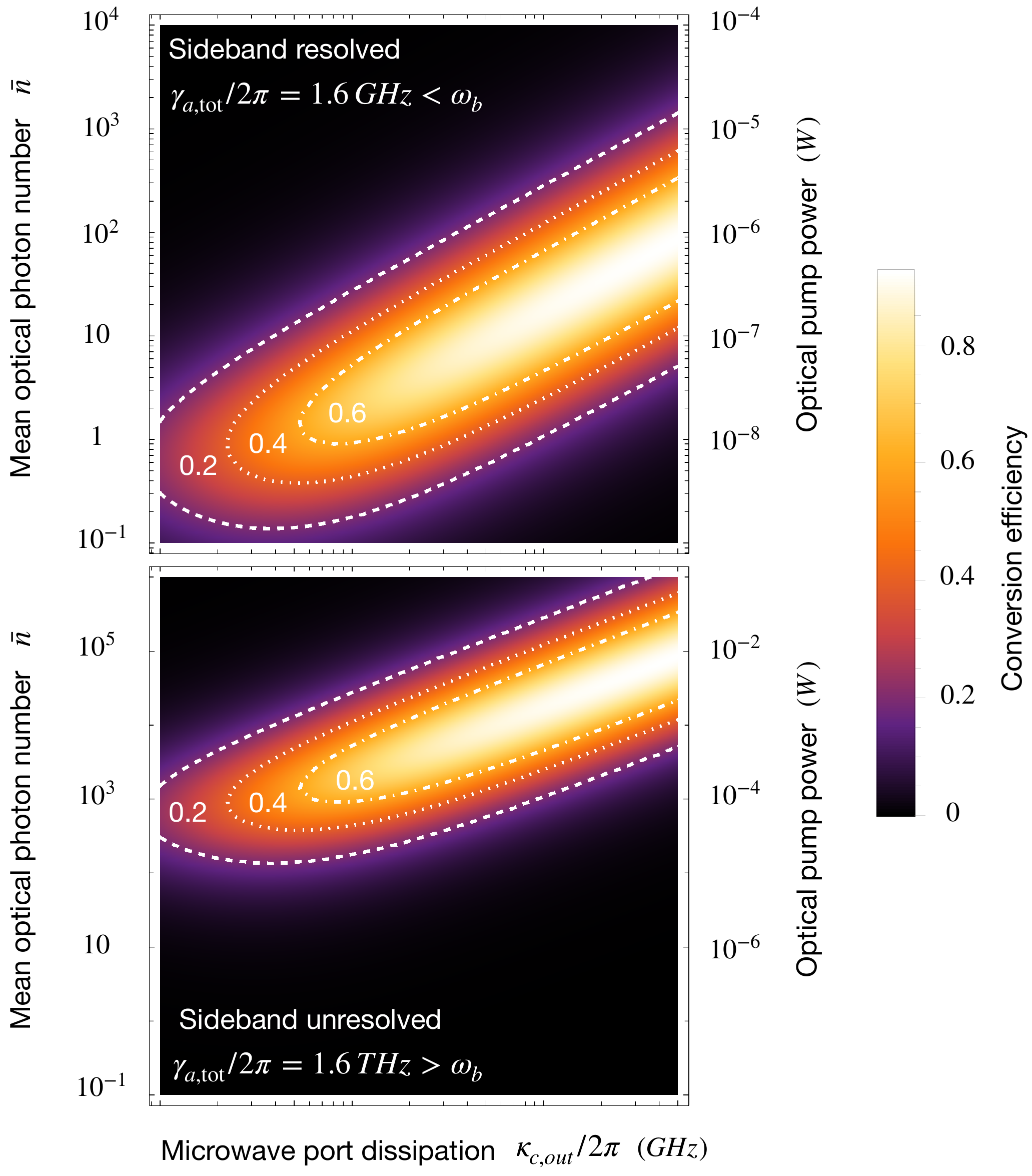}
    \caption{Plot of the conversion efficiency given in Eq. \eqref{eq:protocol_magnomech_convo} as a function of the mean optical photon number per unit volume (optical pump power) and the microwave port dissipation rate. We set $\omega_b = \omega_m =\omega_c = -\Delta_a$. The top plot corresponds to the sideband resolved regime of the optical cavity while the bottom plot is for the unresolved case. The color code denotes the conversion efficiency. Other parameters corresponding to the estimates of Table \ref{tab:values}.}
    \label{fig:protocol_sideband_comp}
\end{figure}
In the case of full mode resonance and red optical detuning ($\omega_b = \omega_m = \omega_c = -\Delta_a $), we plot the conversion efficiency given in Eq. (\ref{eq:protocol_magnomech_convo}) as a function of the mean optical photon number and the microwave port dissipation in Fig. \ref{fig:protocol_sideband_comp}, for two different total optical linewidths $\gamma_{a,\rm{tot}}$. For $\gamma_{a,\rm{tot}}=2\pi\times1.6\,\rm{THz}$ the system is in the sideband unresolved regime of the optical cavity, since the linewidth is larger than the mode frequency of the phonon ($\omega_b=2\pi\times10\,\rm{GHz}$). In this regime it is not important that the optical detuning is chosen precisely. When we compare the two cases, we conclude that both allow to reach the same conversion efficiencies. However, in the sideband unresolved regime we need much stronger optical pump powers to obtain similar results as for the unresolved regime, due to the stronger dissipation.
\subsection{Weak inductive coupling regime}\label{subsec:smallinductive}
In the previous sections, we have assumed a mode overlap between magnon and microwave of $1 \%$, which, based on our estimates presented in the appendix, yields a coupling value of $g_{mc}=180\,\rm{MHz}$. Nevertheless, such overlap is challenging to achieve experimentally, since the system's design would require a balance between a sufficiently small sample volume, which yields a high optomechanical coupling, while maintaining the volume big enough as to achieve a high magnon-microwave cooperativity. In this section we address such design limitation by quantifying the performance of our setup under a magnon-microwave mode overlap of $0.01 \%$, corresponding to $g_{mc}=18\,\rm{MHz}$.  \par
For this value of the coupling, the cooperativity between magnon and microwave is $C_{mc}=13$. Based on our discussion of optimization in terms of cooperativties (following Eq. \eqref{eq:protocol_magnomech_convo}), we still expect high conversion efficiency for sufficiently strong optical pump power. 
\begin{figure*}
    \centering
    \includegraphics[width=1.0\textwidth]{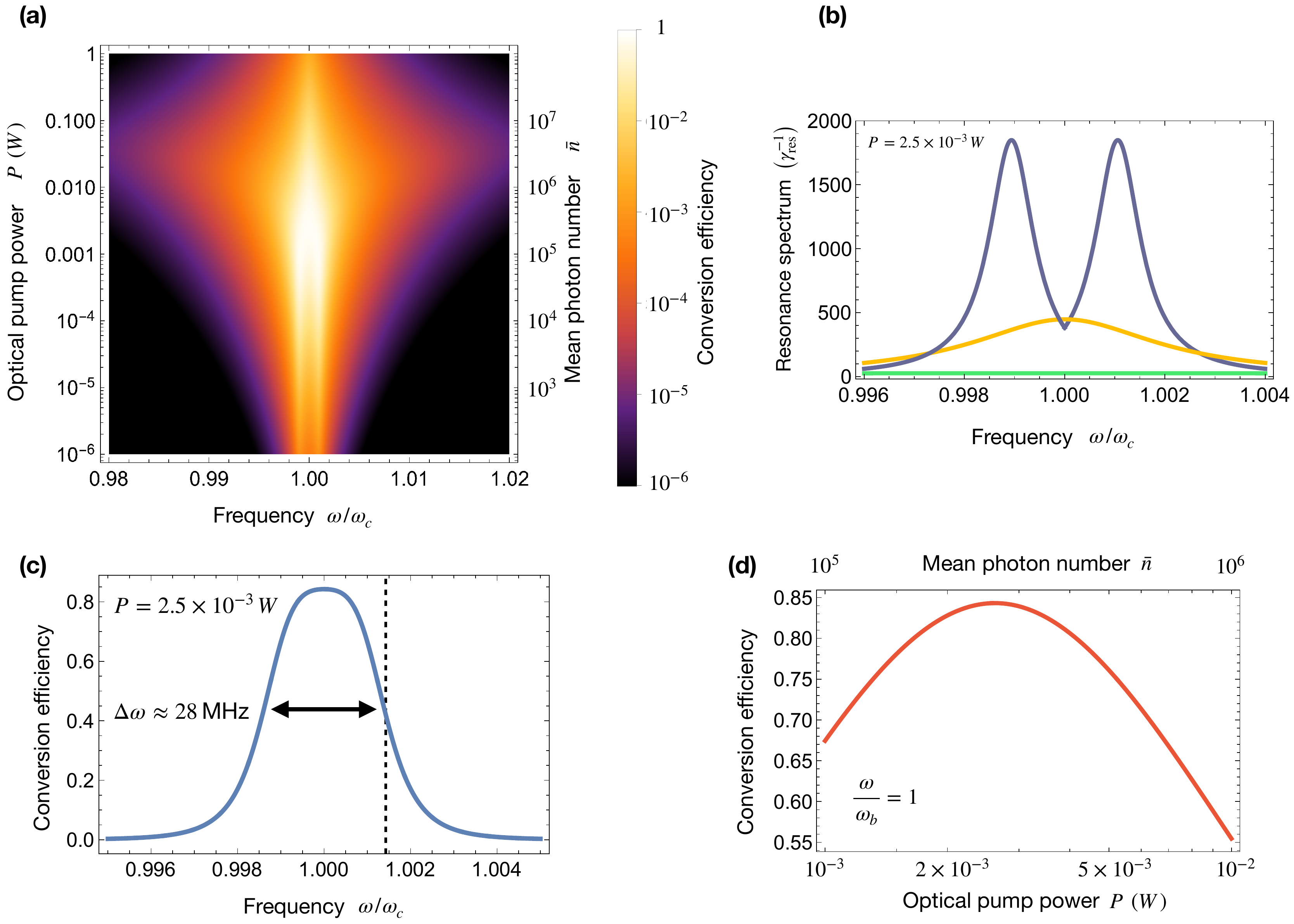}
    \caption{(a) Conversion efficiency as a function of frequency and optical pump power (corresponding mean photon number on right y-axis). The frequency axis is given in units of $\omega_c$, the microwave frequency and we have set $\omega_b = \omega_m =\omega_c = -\Delta_a$. (b) Frequency spectrum of the resonances of the coupled system following Eqs. (\ref{eq:protocol_langevin_frq_a}) to (\ref{eq:protocol_langevin_frq_d}). In this case the blue peaks correspond to the magnon-phonon hybrid modes. Since the system is still in the weak optomechanical coupling regime for the used optical pump power of $P=2.5\times 10^{-3}\,\rm{W}$ ($C_{ab}=28118$, $C_{mc}=13$), the phonon (yellow) and photon (green) are not shifted. (c) Conversion efficiency as a function of frequency for the point of highest efficiency at a pump power of about  $P=2.5\times 10^{-3}\,\rm{W}$. The remarkably large bandwidth at this point is marked with $28\,\rm{MHz}$. (d) Conversion efficiency as a function of optical pump power.}
    \label{fig:altern_inductive}
\end{figure*}
The reduced coupling strength puts the system in the weak inductive coupling regime since $g_{mc}<\gamma_{c, \rm{tot}}$. For small optical pump powers, magnons hybridize with the phonons instead of with microwaves. 
As a consequence, the conversion efficiency as a function of both frequency and optical pump power, depicted in Fig. \ref{fig:altern_inductive} (a), exhibits two peaks shifted by $g_{mb}=10\,\rm{MHz}$ from the resonance point. For increasing pump power the system approaches the strong optomechanical coupling regime. Around $P=2.5\times 10^{-3}\,\rm{W}$ the system reaches the point of largest conversion efficiency (compare Fig. \ref{fig:altern_inductive} (c), (d)) where the phonon linewidth in the resonance spectrum, shown in Fig. \ref{fig:altern_inductive} (b), is strongly broadened, yielding a conversion bandwidth of up to $28\,\rm{MHz}$. Note that for this pump power the system is not yet in the strong optomechanical regime ($G_{ab}\approx 106\,\rm{MHz}<\nicefrac{\gamma_{a,\rm{tot}}}{4}$). For a stronger pump power, the optomechanical coupling surpasses the magnomechanical coupling, and phonons hybridize strongly with optical photons. After transition to the strong optomechanical regime, the hybridization between magnons and phonons is weak, and as a consequence, both the bandwidth and the overall efficiency decrease.\par
Compared to the results discussed in the previous sections, we obtain a much larger conversion bandwidth at a similar conversion efficiency. Therefore, smaller magnon-microwave overlaps can still yield good conversion efficiencies at broad bandwiths. However, the high required optical pump power (and therefore large mean photon number on the order of $10^6$) strongly limits the configuration to application in the classical regime.

\section{Conclusions \label{section:outlook}}

In summary, we derived the efficiency of a two-stage optics-to-microwave conversion system that incorporates hybridized mechanical and magnetic excitations as mediating elements. The conversion efficiency exhibits a spectrum that can be understood in terms of the hybrid modes formed by the strong interaction of the coupled modes. In a setup in which magnons, microwaves and phonons are resonant while the laser drive is red detuned, the efficiency can be expressed solely in terms of the cooperativities, and we have derived a condition for the maximum efficiency as a function of the cooperativities which can be used to determine optimal values for port coupling rates and optical pump power. Furthermore, we discussed the conversion bandwidth for different optical pump powers and the effect of total linewidth of the optical cavity mode. Our results do not take into account any specific geometry and can be used as a starting point for designing a magnomechanical-based microwave to optics converter.\par
We have shown that high conversion between optical and microwave modes is possible by the resonant coupling between phonons and magnons. In regions where the conversion efficiency is large we obtain bandwidths on the same order of magnitude as the coupling rates. This is an advantage over state-of-the-art one-stage conversion setups based on e.g. optomechanics/electromechanics, which are generally limited by their small total linewidth of the vibrational mode. Moreover, the inclusion of magnons adds a desirable tunability to the system and their coupling to phonons overcomes the challenges of weak optomagnonic interactions. \par
The work presented in this article applies to a fully classical framework and application in quantum conversion requires the evaluation of the added noise, which in turns contributes to the total state transduction fidelity. In general we expect that vacuum and thermal noise is added during each step in the conversion chain \cite{hanMicrowaveopticalQuantumFrequency2021}. Since the optical mode frequency is in the THz regime, thermal occupation for optics is expected to be negligible or vanish, even for room temperature, thus contributing only with vacuum noise. The added thermal noise for the microwave is suppressed when the port is strongly overcoupled and the device is operated at sufficiently low temperature. To assure low added quantum noise from the intermediate elements (magnons, phonons) we expect constraints that are potentially stricter than those required for high conversion efficiency \cite{zeuthenFiguresMeritQuantum2020}. Finally it is important to note that even though the conversion efficiency is the same in both directions, the noise adds differently to the optic and microwave ports. \par
A prospective geometry where the proposed system could be implemented would have to fulfil strict requirements in terms of frequency matching, and ensure good overlap and symmetry matching of the participating modes in order to achieve high coupling rates. Patterned magnetic dielectrics at the microscale in the form of an optomechanical crystal can be  promising in this regard \cite{grafDesignOptomagnonicCrystal2021a}. Magnon-phonon coupling in this kind of structures remains to be investigated. Our choice of 1 $\mu m^3$ for the effective magnon mode volume at 1$\%$ mode overlap with the microwave field also requires improvement of microwave resonator design capable to achieve such a small microwave mode volume, for example through a planar geometry. Alternatively, one could build an overall larger geometry, at the expense of a reduced single-particle optomechanical coupling. The drawbacks of both approaches can be partially resolved by increasing the optomechanical cooperativity (see Fig. \ref{fig:appendix_coop}), e.g. via a stronger optical drive. We have discussed the case of a worse mode overlap between magnons and microwaves of only $0.01\%$, which yields comparable results for conversion efficiency with a larger conversion bandwidth. A well optimized device would represent a viable technique for overcoming the challenge of efficient quantum frequency conversion, and pave the way towards long distance quantum communication.
\section*{Acknowledgements}
F.E., V.A.S.V.B. and S.V.K. acknowledge financial support from the Max Planck Society and from the Deutsche Forschungsgemeinschaft (DFG, German Research Foundation) through Project-ID 429529648–TRR 306 QuCoLiMa (“Quantum Cooperativity of Light and Matter”). H.H. acknowledges funding via the Deutsche Forschungsgemeinschaft (DFG, German Research Foundation) under Germany’s Excellence Strategy-EXC-2111-390814868. This work was partially supported by the French Grants ANR-18-CE24-0021 Maestro and ANR-21-CE24-0031 Harmony.
\newpage
\appendix
\newpage
\appendix
\begin{table*}
\caption{\label{tab:values_mat}Table of used values for material specific constants.}
\begin{ruledtabular}
\begin{tabular}{ccc}
 Quantity  &Symbol &Value \\ \hline
 Saturation Magnetization               & $M_S$                             & $140\time10^3 \,\rm{A\cdot m^{-1}}$  \\ 
 
 Magnetoelastic constant (major axes)   & $B_1$                             & $3.48\,\rm{J\cdot m^{-3}}$  \\
 
 Magnetoelastic constant                & $B_2$                             & $6.4\,\rm{J\cdot m^{-3}}$  \\

 Mass density (YIG)                     & $\rho$                            & $5110\,\rm{kg\cdot m^{-3}}$  \\
 
 Spin density (YIG)                     & $\rho_{\rm{S, YIG}}$              & $4.22\times10^{27}\,\rm{m^{-3}}$  \\
 
 Relative Permittivity ($1.2\,\mu \rm{m}$)   & $\varepsilon_r$                   & $5$  \\

 Refractive Index ($1.2\,\mu \rm{m}$)        & $n$                               & $2$  \\
 
 Vacuum Permittivity                    & $\varepsilon_0$                   & $8.85\times10^{-12}\,\rm{{F\cdot m^{-1}}}$  \\ 
 
 Vacuum Permeability                    & $\mu_0$                           & $1.26\times10^{-6}\,\rm{H\cdot m^{-1}}$  \\ 
 
 Mode volume (photon, phonon, magnon)   & $V_{a/b/m}^{\rm{eff}}$            & $1\,\mu \rm{m^3}$  \\ 
\end{tabular}
\end{ruledtabular}
\end{table*}
\section*{Appendix: Coupling derivations and estimates} 
%
%
\subsection{Magnomechanical coupling \label{appendix:magnomech}}
Magnetic properties inside a material are defined by the collective interaction of its atomic magnetic moments. Mechanical deformation alters this collective interaction and gives rise to coupling between mechanical and magnetic degrees of freedom. This magnomechanical interaction is captured by the magnetoelastic anisotropy energy, as it is the only energy that relates the magnetization direction with the crystal axes. Formally, the magnetoelastic anisotropy energy is defined as the difference between the energies required to magnetize a material along an easy axis and a hard axis. 
The magnetoelastic energy is given by \cite{gurevichMagnetizationOscillationsWaves2020}
\begin{eqnarray}
    &&E_{\mathrm{mel}}=\frac{B_{1}}{M_{S}^{2}}\int d^{3}r\left[M_{x}^{2}\sigma_{xx}+M_{y}^{2}\sigma_{yy}+M_{z}^{2}\sigma_{zz}\right]\\ &&+\frac{2B_{2}}{M_{S}^{2}}\int d^{3}r\left[M_{x}M_{y}\sigma_{xy}+M_{y}M_{z}\sigma_{yz}+M_{z}M_{x}\sigma_{zx}\right].\nonumber
    \label{eq:appendix_magnetoelastic_energy}
\end{eqnarray}
We quantize this energy in terms of both spin waves (magnons) and mechanical modes (phonons). Let the total magnetization be defined as a static magnetic groundstate plus a dynamic deviation on top, so $\textbf{m}\left(\textbf{r},t\right)=m_S\left(\textbf{r}\right)+\delta m\left(\textbf{r},t\right)$. Note that here we introduced the normalized magnetic field components $\textbf{m}=\nicefrac{\textbf{M}}{\textbf{M}_S}$. The temporally dynamic deviation is quantized via bosonic annihilation and creation operators 
\begin{equation}
    \delta\mathbf{\hat{m}}\left(\mathbf{r},t\right)=\underset{{\scriptstyle \alpha}}{\sum}\delta\mathbf{m_{\alpha}}\left(\mathbf{r}\right)e^{-i\omega_{\alpha}t}\hat{m}_{\alpha}+\delta\mathbf{m_{\alpha}^{\ast}}\left(\mathbf{r}\right)e^{i\omega_{\alpha}t}\hat{m}_{\alpha}^{\dagger},
\end{equation}
where we are summing over multiple modes $\alpha$. The mechanical modes are quantized in the same way
\begin{equation}
    \hat{\mathbf{u}}(\mathbf{r},t)=\underset{{\scriptstyle \eta}}{\sum}\mathbf{u}_{\eta}(\mathbf{r})\hat{b}_{\eta}e^{-i\omega_{\eta}t}+\mathbf{u}_{\eta}^{\ast}(\mathbf{r})\hat{b}_{\eta}^{\dagger}e^{i\omega_{\eta}t},
    \label{eq:Appendix_phonon_quant}
\end{equation}
where we again sum over multiple modes $\eta$. Note that in the definition of the magnetoelastic energy (Eq. \ref{eq:appendix_magnetoelastic_energy}) we have elements of the symmetric strain tensor
\begin{equation}
    \sigma_{ij}=\frac12\left(\frac{\partial u_i}{\partial j}+\frac{\partial u_j}{\partial i}\right),
\end{equation}
where $i,j \in \left(x,y,z\right)$, so we have the spatial derivatives of the mechanical mode profiles entering the coupling. The quantized magnomechanical coupling Hamiltonian reads
\begin{align}
    \hat{H}_{\mathrm{ME}}=&\underset{{\scriptstyle \alpha\beta\eta }}{\sum}G_{\alpha\eta}\hat{m}_{\alpha}\left(\hat{b}_{\eta}+\hat{b}_{\eta}^{\dagger}\right)+T_{\alpha\beta\eta}\hat{m}_{\alpha}\hat{m}_{\beta}\left(\hat{b}_{\eta}+\hat{b}_{\eta}^{\dagger}\right)\nonumber \\& +P_{\alpha\beta\eta}\hat{m}_{\beta}^{\dagger}\hat{m}_{\alpha}\left(\hat{b}_{\eta}+\hat{b}_{\eta}^{\dagger}\right)+h.c.,
\end{align}
with the coupling strengths defined as the following overlap integrals
\begin{eqnarray}
    G_{\alpha\eta}&&=2B_{1}\int d^{3}r\sum_{i}\:m_{S}^{i}\delta m_{\alpha}^{i}\partial_i\,u_{\eta}^{i}+\nonumber\\&&B_{2} \int d^{3}r \sum_{ij}\left(m_{S}^{i}\delta m_{\alpha}^{j}+m_{S}^{j}\delta m_{\alpha}^{i}\right)\times \nonumber\\
    &&\quad \quad \quad \quad \quad \quad \quad \left(\partial_{j}\,u_{\eta}^{i}+\partial_{i}\,u_{\eta}^{j}\right)\\
    T_{\alpha\beta\eta}&&=B_{1}\int d^{3}r\sum_{i}\:\delta m_{\alpha}^{i}\delta m_{\beta}^{i}\partial_i\,u_{\eta}^{i}+\nonumber\\ && B_{2} \int d^{3}r \sum_{ij}\delta m_{\alpha}^{i}\delta
    m_{\beta}^{j}\left(\partial_{j}\,u_{\eta}^{i}+\partial_{i}\,u_{\eta}^{j}\right),\\
    P_{\alpha\beta\eta}&&=B_{1}\int d^{3}r\sum_{i}\:\left(\delta m_{\alpha}^{i}\right)^{\ast}\delta m_{\beta}^{i}\partial_i\,u_{\eta}^{i}+\nonumber \\&&B_{2} \int d^{3}r \sum_{ij}\left(\delta m_{\alpha}^{i}\right)^{\ast}\delta m_{\beta}^{j}\left(\partial_{j}\,u_{\eta}^{i}+\partial_{i}\,u_{\eta}^{j}\right).
\end{eqnarray}
The couplings are normalized by the total quasi-particle numbers of both magnons and phonons. The phonon number is calculated using the mechanical energy contained in given mode $\eta$
\begin{equation}
    N_{\textrm{phonon},\eta}=\frac{2\omega_{\eta}^2 \rho V_{\textrm{phonon},\eta}^{\mathrm{eff}}\cdot\max\left[\left|\mathbf{u}\left(\mathbf{r}\right)\right|^{2}\right]}{\hbar \omega_{\eta}},
    \label{eq:Appendix_norm_phonon}
\end{equation}
where we have used the definition of the kinetic energy. The magnetic part is normalized by the average number of spins in a given magnetic mode $\alpha$
\begin{equation}
    N_{\mathrm{mag},\alpha}=\frac{M_{S}V_{\mathrm{mag},\alpha}^{\mathrm{eff}}}{g\mu_{B}}.
    \label{eq:Appendix_norm_magnon}
\end{equation}
For linear magnomechanical interaction $g_{mb}\hat{m}\left(\hat{b}+\hat{b}^{\dagger}\right)+h.c.$ the coupling strength is given by
\begin{eqnarray}
    g_{mb}^{\alpha \beta}&&= 2B_1 \int d^3 r \sum_{i}\, m_0 ^i \delta m_{\alpha} ^i \sigma_{\beta} ^{ii}\nonumber \\
    &&+ B_2 \int d^3 r \sum_{i, j}\, \left( m_0 ^i \delta m_{\alpha} ^j + m_0 ^j \delta m_{\alpha} ^i\right)\sigma_{\beta}^{ij},
    \label{eq:model_magnomechCoup}
\end{eqnarray}
where $m_0$ denotes the magnetic ground state, $\delta m$ the magnetic mode and $\sigma^{ij}$ the symmetric strain tensor. $B_{1,2}$ are the material specific magnetoelastic constants and we have modes $\left(\alpha, \beta\right)$ and spatial indices $\left(i,j\right)=(x,y,z)$. Eq. (\ref{eq:model_magnomechCoup}) represents an overlap integral over the crystal volume. Normalizing by the total magnon and phonon numbers inside given respective modes, we can estimate the single mode coupling for perfect mode overlap, so $V_m^{\rm{eff}} = V_b^{\rm{eff}} = V$. In this case, the integral is also equal to $V$ and the coupling is independent of the mode volumes, so  
\begin{equation}
    g_{mb} \approx (2B_1 + B_2) \sqrt{\frac{\hbar g \mu_B}{2\omega_b \rho \max[\textbf{u}]^2M_S}}.
\end{equation}
Here, $g$ is the gyromagnetic ratio, $\mu_B$ the Bohr magneton, $\rho$ the mass density and $M_S$ the saturation magnetization. The volumes $V_m$ and $V_b$ correspond to spatial mode volumes of the magnetic and mechanical mode respectively. Furthermore, $\max[\textbf{u}]$ is defined to be the maximum value of the displacement field corresponding to the zero point fluctuation $x_{\rm{ZPF},\eta}$ of a given mechanical mode $\alpha$. Using material specific constants for Yttrium Iron Garnet (YIG) and assuming perfect overlap of mechanical and magnetic mode we obtain coupling rates of $2.5\,\textsc{MHz}$ and $12\,\textsc{MHz}$ for the terms proportional to $B_1$ and $B_2$, respectively. The mechanical mode frequency was chosen to $\omega_b=2\pi\times10\,\textsc{GHz}$. \par 
\subsection{Photoelastic coupling}
The interaction between phonons and optical photons arises mostly from the photoelastic effect \cite{nelsonNewSymmetryAcoustoOptic1970}. However, the total coupling strength is also affected by the deformation of the boundary. This contribution is usually on the order of $10\%$ of that of the photoelastic effect and is difficult to estimate since it has to be evaluated for a specific geometry. For our estimate we will neglect this contribution. A mechanical mode deforms the optical cavity therefore altering its resonance frequency $\omega_{0}$. This requires that a potential coupling device serves both as mechanical resonator and optical cavity. We introduce a a small perturbation in the permittivity tensor of the material via $\bar{\bar{\varepsilon}}\left(\mathbf{r}\right)=\bar{\bar{\varepsilon}}_{0}\left(\mathbf{r}\right)+\bar{\bar{\delta\varepsilon}}\left(\mathbf{r}\right)$. To first order, we obtain the shift of the resonance frequency for a given optical mode profile $\mathbf{e}\left(\mathbf{r}\right)$ and write the photoelastic overlap integral as \cite{aspelmeyerCavityOptomechanicsNano2014a}
\begin{eqnarray}
    E_{\textrm{photoelastic}}&&=-\varepsilon_{0}n^{4}\int d^{3}r\,\mathbf{e}\cdot\bar{\bar{\delta\varepsilon}}\cdot\mathbf{e}\nonumber\\
	&&=-\varepsilon_{0}n^{4}\int d^{3}r \Big[\,2Re\left\{ E_{i}^{*}E_{j}\right\} p_{44}u_{ij}+\nonumber\\
	&&\left|E_{i}\right|^{2}\left(p_{11}\sigma_{ii}+p_{12}\left(\sigma_{jj}+\sigma_{kk}\right)\right)\Big].
\end{eqnarray}
The optical modes are quantized via bosonic operators
\begin{equation}
    \mathbf{E}\left(\mathbf{r},t\right)=\underset{\mu}{\sum}=\mathbf{E}_{\mu}\left(\mathbf{r}\right)e^{-i\omega_{\mu}t}\hat{a}_{\mu}+\mathbf{E}_{\mu}^{*}\left(\mathbf{r}\right)e^{i\omega_{\mu}t}\hat{a}_{\mu}^{\dagger}.
    \label{eq:Appendix_quant_em_modes}
\end{equation}
Together with the quantization of the mechanical mode (Eq. \ref{eq:Appendix_phonon_quant}), we obtain the quantized photoelastic interaction Hamiltonian as
\begin{equation}
    \hat{H}_{\mathrm{PE}}=\underset{\mu\nu\eta}{\sum}g_{\mu\nu\eta}\hat{a}_{\mu}^{\dagger}\hat{a}_{\nu}\left(\hat{b}_{\eta}^{\dagger}+\hat{b}_{\eta}\right),
\end{equation}
with coupling strength defined as the overlap integral
\begin{eqnarray}
    g_{\mu\nu\eta}&&=-\varepsilon_{0}n^{4}\int d^{3}r\Big[\sum_{ij}\,\left(E_{\mu}^{i}\right)^{*}E_{\nu}^{j}\sigma_{\eta}^{ij}p_{44}\nonumber \\ &&+\sum_{ijk}\left|E_{\mu}^{i}\right|^{2}\left(\sigma_{\eta}^{ii}p_{11}+\left(\sigma_{\eta}^{jj}+\sigma_{\eta}^{kk}\right)p_{12}\right)\Big].
\end{eqnarray}
where we have modes $\left(\mu, \nu, \eta \right)$ and spatial indices $\left(i,j,k\right)=(x,y,z)$. This integral is again normalized to coupling per phonon (Eq. \ref{eq:Appendix_norm_phonon}) and photon. The number of photon contained in an optical mode $\alpha$ is calculated via
\begin{equation}
    N_{\textrm{photon},\mu}=\frac{2V_{\textrm{photon},\mu}^{\mathrm{eff}}\cdot\max\left[\mathbf{E}^{*}\bar{\bar{\varepsilon}}\mathbf{E}\right]}{\hbar\omega_{\textrm{photon},\mu}},
    \label{eq:Appendix_norm_photon}
\end{equation}
where we have used the definition of the total electric energy. The normalized coupling for photoelastic interactions neglecting the boundary contributions reads
\begin{eqnarray}
    &&g_{ab}^{\mu\nu\eta} = -\frac{n^4}{\epsilon_r}\frac{\hbar \omega_a}{2V_a}\sqrt{\frac{\hbar}{2\omega_b \rho V_b \max[\textbf{u}]^2}}\nonumber \times \\ 
    &&\times\int d^3 r \Big[ \sum_{i j }\, \left(e_{\mu}^i\right)^{\ast} e_{\nu}^j \sigma_{\eta}^{ij}p_{44} +  \nonumber\\ &&\sum_{i j k } \mid e_{\mu}^i\mid^2\left(\sigma_{\eta}^{ii}p_{11}+\left(\sigma_{\eta}^{jj}+\sigma_{\eta}^{kk}\right)p_{12}\right)\Big], 
    \label{eq:optomech_coup_est}
\end{eqnarray}
where $n$ is the refractive index, $\epsilon_r$ the relative permittivity, $e$ the electric field components and $p$ the elements of the material specific photoelastic tensor. For perfect mode overlap, the single mode coupling approximates to $0.2\,\textsc{MHz}$ for material specific constant of YIG. We consider a phonon mode frequency of $\omega_b=2\pi\times10\,\textsc{GHz}$ and an optical mode frequency of $\omega_a=2\pi\times200\,\textsc{THz}$. 
\subsection{Magnon-microwave coupling}
The linear interaction between a microwave cavity field and the magnetic excitations is mediated by magnetic dipole interaction, captured by the energy term
\begin{equation}
    E_{mag}=\frac12 \mu_0 \int d^3r \, \textbf{M}\cdot \textbf{H}.
\end{equation}
The magnetic energy can be quantized in terms of cavity modes. Following Eq. \eqref{eq:Appendix_quant_em_modes}, but for a magnetic field profile, we have 
\begin{equation}
    \mathbf{H}\left(\mathbf{r},t\right)=\underset{\eta}{\sum}=\mathbf{H}_{\eta}\left(\mathbf{r}\right)e^{-i\omega_{\eta}t}\hat{c}_{\eta}+\mathbf{H}_{\eta}^{*}\left(\mathbf{r}\right)e^{i\omega_{\eta}t}\hat{c}_{\eta}^{\dagger}.
    \label{eq:Appendix_quant_B_modes}
\end{equation}
and normalized to a coupling per microwave photon via
\begin{equation}
    N_{\textrm{mw-photon},\mu}=\frac{2V_{\textrm{mw-photon},\mu}^{\mathrm{eff}}\cdot\max\left[\mathbf{H}^{*}\bar{\bar{\mu}}\mathbf{H}\right]}{\hbar\omega_{\textrm{mw-photon},\mu}},
    \label{eq:Appendix_norm_mwphoton}
\end{equation}
and magnon using Eq. \eqref{eq:Appendix_norm_magnon}. The normalized overlap integral reads
\begin{equation}
    g_{mc}^{\alpha \eta} = \frac{1}{2}\sqrt{\frac{\mu_0 \hbar \omega_c g \mu_B M_S}{2V_m V_c}} \int d^3 r \, \delta \textbf{m}_{\alpha} \cdot \textbf{h}_{\eta},
    \label{eq:model_microwaveMagnonCoup}
\end{equation}
where a magnetic mode $\delta \textbf{m}$ couples to the magnetic part $\textbf{h}$ of an electromagnetic microwave field. For perfect mode overlap, the single mode coupling strength is evaluated to 
\begin{equation}
    g_{cm} \approx \frac{g\mu_{B}}{2h} \sqrt{{\mu_{0}\hbar\omega_{c} \rho_S }},
\end{equation}
where $\rho_S$ is the material specific spin density. The coupling can be evaluated to $1.85\,\textsc{GHz}$ for a microwave mode frequency of $\omega_c=2\pi\times10\,\textsc{GHz}$. Perfect overlap is challenging to achieve experimentally, therefore we have considered an overlap factor of $1 \%$, such that $V_m = 0.01V_c$, yielding a coupling rate of $180\,\text{MHz}$. We further discuss the performance of the system under a worse mode overlap of $0.01 \%$ ($V_m = 0.0001V_c$), for which the coupling is $18\,\text{MHz}$.
\bibliography{DraftOMM}
\end{document}